\documentclass[prd,aps,a4paper,preprint,preprintnumbers,nofootinbib]{revtex4-1}
\usepackage[a4paper, hdivide={1.91cm,,1.165cm}, vdivide={1.83cm,,3.2cm}]{geometry}

\usepackage{amsmath,amssymb}
\usepackage{graphicx,multirow}
\usepackage{tabularx}
\usepackage[export]{adjustbox}
\usepackage[hyperfootnotes=false]{hyperref}
\usepackage[usenames, dvipsnames]{color}

\newcommand{\ii}{\ensuremath{\text{i}}}
\newcommand{\dd}{\ensuremath{\text{d}}}

\newcommand{\beq}{\begin{eqnarray}}% can be used as {equation} or  {eqnarray}
\newcommand{\eeq}{\end{eqnarray}}

\begin{document}

%%%%%%%%%%%%%%%%%%%%%%%%%%%%%%%%%

\preprint{UCI-TR-2021-17}

\title{Proca Q-balls and Q-shells}

\author{Julian~Heeck}
\email{heeck@virginia.edu}
\affiliation{Department of Physics, University of Virginia,
Charlottesville, Virginia 22904-4714, USA}

\author{Arvind~Rajaraman}
\email{arajaram@uci.edu}
\affiliation{Department of Physics and Astronomy, 
University of California, Irvine, CA 92697-4575, USA
}

\author{Rebecca~Riley}
\email{rebecca.riley@uci.edu}
\affiliation{Department of Physics and Astronomy, 
University of California, Irvine, CA 92697-4575, USA
}

\author{Christopher~B.~Verhaaren}
\email{verhaaren@physics.byu.edu}
\affiliation{Department of Physics and Astronomy, Brigham Young University, Provo, UT, 84602, USA
}

\date{\today}

\begin{abstract}
Non-topological solitons such as Q-balls and Q-shells have been studied for scalar fields invariant under global and gauged $U(1)$ symmetries. We generalize this framework to include a Proca mass for the gauge boson, which can arise either from spontaneous symmetry breaking or via the St\"uckelberg mechanism. A heavy (light) gauge boson leads to solitons reminiscent of the global (gauged) case, but for intermediate values these Proca solitons exhibit completely novel features such as disconnected regions of viable parameter space and Q-shells with unbounded radius. We provide numerical solutions and excellent analytic approximations for both Proca Q-balls and Q-shells. These allow us to not only demonstrate the novel features numerically, but also understand and predict their origin analytically.
\end{abstract}

%%%%%%%%%%%%%%%%%%%%%%%%%%%%%%%%%
\maketitle

\tableofcontents
%%%%%%%%%%%

\section{Introduction\label{s.intro}}

Scalar field theories with a conserved Noether charge $Q$ can support non-topological soliton solutions~\cite{Lee:1991ax,Nugaev:2019vru}, which we refer to simply as solitons. The simplest examples are Q-balls from $U(1)$-invariant complex scalars~\cite{Coleman:1985ki,Heeck:2020bau}; the required conserved symmetry could also be a \emph{local} symmetry, which leads to gauged solitons in the form of Q-balls~\cite{Lee:1988ag,Gulamov:2015fya,Gulamov:2013cra,Heeck:2021zvk} and Q-shells~\cite{Arodz:2008nm,Tamaki:2014oha,Heeck:2021gam}.

The additional generalization to \emph{massive} gauge bosons has garnered scant attention in the literature (see, however, Refs.~\cite{Ishihara:2018rxg,Ishihara:2019gim,Forgacs:2020vcy,Ishihara:2021iag}) because it is a significantly more difficult system to solve, even numerically.
However, given the prevalence of massive gauge bosons in the Standard Model and its extensions, it is of great phenomenological interest to investigate the effect of gauge boson masses.\footnote{We should mention that solitons in models of self-interacting massive gauge bosons without any scalars have been discussed in Refs.~\cite{Loginov:2015rya,SalazarLandea:2016bys,Brihaye:2017inn,Herdeiro:2020jzx} but have no relation to our study.}
Clearly, a very light gauge boson leads to solitons resembling the known gauged ones, while very heavy gauge bosons lead to solitons reminiscent of global Q-balls. In addition to quantifying the previous statement, we show below that gauge boson masses in the intermediate range exhibit novel effects. 

We begin, in Sec.~\ref{s.setup}, by defining the theory and introducing Q-balls and Q-shells. The mechanical analogy of a particle rolling in a two-dimensional potential is introduced and the different soliton types are associated with different trajectories along this potential. This language provides familiar terminology and intuition for understanding the various solitons considered. The numerical methods employed to determine the exact soliton solutions are also introduced.

Following this discussion, in Sec.~\ref{sec:balls} we focus on Proca Q-balls, deriving excellent approximate analytic formulae for the scalar and gauge field profiles. We also develop a mapping between the solution space of global Q-balls and Proca Q-balls. This mapping provides an understanding of how Proca Q-balls interpolate between global and gauged Q-balls as the mass of the gauge field changes. It also predicts new features in the Proca Q-ball solutions space, such Q-balls with large minimum charge and gaps in continuity between branches of solutions. These predictions are then compared to exact numerical results and found to be in superb agreement.

We then turn to Q-shells in Sec.~\ref{sec:shells}. Similar to the Q-ball case, we develop impressive analytical approximations of the Q-shell fields. These are used in conjunction with the
mechanical analogy to predict the solutions space of Proca Q-shells. We find not only Q-shells that match onto gauged Q-shells as the gauge boson mass is taken to zero, but also Q-shells without a massless analogue. These predictions are compared to exact numerical results and shown to agree. Following this discussion we conclude in Sec.~\ref{s.con} and include some technical derivations regarding general Proca solitons in Appendix~\ref{app:energies}.

\section{Framework\label{s.setup}}

Our starting point for Proca solitons is the following Lagrange density of a complex scalar field~$\phi$ charged under a $U(1)$~gauge symmetry,
\begin{equation}
\mathcal{L}=\left|D_\mu\phi \right|^2-U(|\phi|)-\frac14 F_{\mu\nu}F^{\mu\nu}+\frac{m_A^2}{2}A_\mu A^\mu \,,
\end{equation}
where $U(|\phi|)$ is a $U(1)$-invariant scalar potential, $D_\mu=\partial_\mu-\ii e A_\mu$ is the gauge covariant derivative, and $F_{\mu\nu}=\partial_\mu A_\nu-\partial_\nu A_\mu$ is the gauge field strength tensor. The (positive) parameter $e$ is the gauge coupling normalized so that $\phi$ has charge one. 
We include a mass $m_A$ for the gauge field that can have two origins: 
\begin{enumerate}
\item[i)] it can be considered a St\"uckelberg mass~\cite{Stueckelberg:1938hvi,Ruegg:2003ps}, in which case the $U(1)$ gauge symmetry is unbroken albeit fixed in $\mathcal{L}$, while the underlying \emph{global} $U(1)$ symmetry $\phi (t,\vec x)\to e^{\ii \theta} \phi (t,\vec x)$ is manifestly conserved; 
\item[ii)]  it can originate from spontaneous symmetry breaking by a scalar $\psi$ that is either very heavy or has negligible couplings to $\phi$ in order to be irrelevant for the soliton dynamics. In the latter case the full Lagrangian $\mathcal{L}(\phi,\psi)$ must feature one global and one gauged $U(1)$ symmetry, only the latter of which is broken by $\langle \psi\rangle$.
\end{enumerate}
Assuming the field $\psi$ in ii) can be neglected, the phenomenology of $\mathcal{L}$ is the same no matter the origin of $m_A$. In the following we simply refer to the massive gauge boson $A$ as a Proca field and the corresponding solitons as Proca solitons. These are to be distinguished from global solitons (the limiting case $e\to 0$ or $m_A\to \infty$) and gauged solitons (the limiting case $m_A\to 0$).

We expect ground-state soliton solutions to be spherically symmetric~\cite{Coleman:1985ki} and make the following ansatz for the fields,
\begin{align}
\phi (t,\vec x)=\frac{\phi_0}{\sqrt{2}}f(r) e^{\ii\omega t}\,, \quad
 A_0 (t,\vec x)= \phi_0 A(r)\,, \quad
 A_{1,2,3}(t,\vec x) =0\,,
\label{e.fields}
\end{align}
with a constant frequency $\omega$. The dimensionful scale constant $\phi_0$ is determined by the scalar potential.
We have left $U(|\phi|)$ largely unspecified, but there are a few qualities that must be present. Since we do not want to break the $U(1)$ symmetry, at least not through $\phi$, we require $\langle\phi\rangle=0$ in the vacuum, which implies that $f=0$ in the vacuum. Then, we choose the potential energy to be zero at the vacuum by $U(0)=0$ and enforce that the vacuum is a stable minimum of the potential by 
\begin{equation}
\left.\frac{\dd U}{\dd |\phi|}\right|_{\phi=0}=0~, \ \ \ \ \left.\frac{\dd ^2U}{\dd \phi\, \dd\phi^\ast}\right|_{\phi=0}=m_\phi^2~,
\end{equation}
where $m_\phi$ is the mass of the complex scalar. With all this in mind, Coleman~\cite{Coleman:1985ki} showed that \emph{global} Q-balls exist when the function $U(|\phi|)/|\phi|^2$ has a minimum at $|\phi |=\phi_0/\sqrt{2}$ with $0<\phi_0<\infty$ such that 
\begin{equation}
0\leq\sqrt{\frac{2U(\phi_0/\sqrt{2})}{\phi_0^2}}\equiv \omega_0< \omega<m_\phi~.\label{e.Omega0}
\end{equation}
Note that we can use the global $U(1)$ symmetry to make $\phi_0$ real and positive without loss of generality.
The gauged and Proca solitons have a somewhat modified region of validity for $\omega$ that is discussed below, but at the very least it is bounded from above by $\omega \leq m_\phi$.

It proves convenient to introduce the dimensionless quantities defined by
\begin{align}
\rho &\equiv r\sqrt{m_\phi^2-\omega_0^2}\,, & 
\Omega &\equiv\frac{\omega}{\sqrt{m_\phi^2-\omega_0^2}}\,, &
\Omega_0 &\equiv\frac{\omega_0}{\sqrt{m_\phi^2-\omega_0^2}}\,,\\
\Phi_0 &\equiv\frac{\phi_0}{\sqrt{m_\phi^2-\omega_0^2}}\,, & 
\alpha &\equiv e\Phi_0 \,, &
M &\equiv\frac{m_A}{\sqrt{m_\phi^2-\omega_0^2}}\,.
\label{eq:rescaling}
\end{align}
In addition, we often use the parameter  $\kappa \equiv \sqrt{\Omega^2 -\Omega_0^2}\in (0,1]$ instead of $\Omega$.
With these definitions and the field ansatz from Eq.~\eqref{e.fields} the Lagrangian takes the form
\begin{equation}
L=4\pi\Phi_0^2 \sqrt{m^2_\phi-\omega_0^2}\int \dd\rho\,\rho^2\left\{ -\frac12f^{\prime 2}+\frac{1}{2}A^{\prime2}+\frac{1}{2}f^2\left(\Omega-\alpha A \right)^2-\frac{U(f)}{\Phi_0^2(m_\phi^2-\omega_0^2)^2}+\frac{M^2}{2}A^2\right\},\label{e.massesLag}
\end{equation}
with primes denoting derivatives with respect to $\rho$.
We also define a (dimensionless) potential for the two dynamical fields:
 \beq
 V(f,A)=\frac{1}{2}f^2\left(\Omega-\alpha A \right)^2-\frac{U(f)}{\Phi_0^2(m_\phi^2-\omega_0^2)^2}+\frac{M^2}{2}A^2\,.
 \label{eq:potentialV}
 \eeq
 The Euler--Lagrange equations pertaining to the Lagrangian in Eq.~\eqref{e.massesLag} are
\begin{align}
f'' + \frac{2}{\rho} f' &= -\frac{\partial V}{\partial f} = \frac{1}{\Phi_0^2(m_\phi^2-\omega_0^2)^2}\frac{\dd U}{\dd f}-\left(\alpha A-\Omega\right)^2f~,\label{e.feq}\\
A'' + \frac{2}{\rho} A' &= +\frac{\partial V}{\partial A} = \alpha f^2(\alpha A-\Omega)+M^2A~.\label{e.Aeq}
\end{align}
The boundary conditions for localized solitons are similar to those for gauged Q-balls~\cite{Lee:1988ag}:
\begin{equation}
\lim_{\rho\to0}f'= \lim_{\rho\to\infty}f =  \lim_{\rho\to0}A'= \lim_{\rho\to\infty}A =0 \,.
\end{equation}

The conserved charge $Q$ is defined in the usual way as the integral over the time component of the scalar current~\cite{Lee:1988ag}, which can be expressed as
\begin{align}
Q&=4\pi \Phi_0^2\int \dd\rho\,\rho^2f^2\left(\Omega-\alpha A \right) .\label{e.Qdef}
\end{align}
Finally, the soliton energy is obtained from the Hamiltonian
\begin{align}
E & =4\pi\Phi_0^2 \sqrt{m_\phi^2-\omega_0^2} \int \dd\rho\,\rho^2\left\{ \frac12f^{\prime 2}+\frac{1}{2}A^{\prime2}+\frac{1}{2}f^2\left(\Omega-\alpha A \right)^2+\frac{U(f)}{\Phi_0^2(m_\phi^2-\omega_0^2)^2}+\frac{M^2}{2}A^2\right\}\label{e.Eint}\\
&= \sqrt{m_\phi^2-\omega_0^2} \,\Omega Q+\frac{4\pi\Phi_0^2 \sqrt{m_\phi^2-\omega_0^2}}{3}\int \dd\rho\,\rho^2\left(f^{\prime 2}-A^{\prime2} \right) .
\label{e.EnoU}
\end{align}
The second equation is derived in App.~\ref{app:energies} together with the proof that the popular soliton equation
\beq
\frac{\dd E}{\dd\omega}=\omega\frac{\dd Q}{\dd\omega}~,\label{e.dedw}
\eeq
continues to hold even in the presence of a gauge boson mass.

Aside from the novel inclusion of the gauge boson mass $M$, the above essentially reviews the known gauged soliton formulae following the conventions of Ref.~\cite{Heeck:2021zvk}. It bears repeating, however, that no exact solutions to the coupled differential equations Eqs.~\eqref{e.feq} and~\eqref{e.Aeq} are known. Even numerical solutions are tedious to obtain, especially upon including the gauge boson mass $M$. Therefore, in what follows, we motivate approximate analytical solutions that also serve as seed functions for numerical finite-element solutions.

Most of our discussion employs the most generic $U(1)$-symmetric, sextic potential studied already in Refs.~\cite{Heeck:2020bau,Heeck:2021zvk,Heeck:2021gam}, which can be conveniently parametrized as
\begin{align}
U(f) = \Phi_0^2 (m_\phi^2 - \omega_0^2)^2 \, \frac{f^2}{2} \left[ (1-f^2)^2 +\Omega_0^2 \right] ,
\label{eq:sextic_potential}
\end{align}
leading to the effective potential
\beq
V(f,A)=\frac{1}{2}f^2\left[\kappa^2+\alpha A(\alpha A-2\Omega)-\left(1-f^2\right)^2 \right]+\frac{M^2}{2}A^2 \,.
\eeq

\subsection{The Potential}

We can understand much of the dynamics of gauged Q-balls by considering the potential $V$ of Eq.~\eqref{eq:potentialV}. If we neglect the friction terms we can write the equations of motion as simply
\beq
f''+\frac{\partial V}{\partial f}=0, \ \ \ \ A''-\frac{\partial V}{\partial A}=0~.
\eeq
Note then that we can define the quantity
\beq
\mathcal{E}=\frac12f^{\prime2}-\frac12A^{\prime2}+V(f,A)~,
\eeq
and find that it is conserved:
\beq
\frac{\dd\mathcal{E}}{\dd\rho}=f'\left(f''+\frac{\partial V}{\partial f} \right)-A'\left(A''-\frac{\partial V}{\partial A} \right)=0~.
\eeq
Of course, in general this quantity is not conserved and we see immediately that
\beq
\frac{\dd\mathcal{E}}{\dd\rho}=-\frac{2}{\rho}\left( f^{\prime2}-A^{\prime2}\right)~.
\eeq
However, we can integrate this quantity to find
\beq
V(f(0),A(0))=2\int_0^\infty\frac{\dd \rho}{\rho}\left( f^{\prime2}-A^{\prime2}\right) ,\label{e.WorkFric}
\eeq
where we have used that $f(\infty)=A(\infty)=0$ and that the derivatives of $f$ and $A$ vanish at both boundaries. Of course, the integration can also be taken over any finite range. For instance, in determining the Q-shell radii we consider integrating from zero to the inner radius and separately from the outer radius to infinity. 

The first term in the friction integral~\eqref{e.WorkFric} appears to be the energy lost due to friction as the $f$ field rolls down the potential to the maximum at $f=0$. It would be helpful to have a similar mechanics intuition for $A$. Notice that $A$'s dynamics are determined by $-V$, instead of $V$. Thus, we have that $A$ also rolls from rest down the slope of $-V$ and this also produces a friction term, but with the opposite sign.

Much of the Q-ball and Q-shell profiles can be extracted by considering the shape of the potential as if it were determining the motion of a particle in two dimensions. 
For constant $f$, the potential $V$ for $A$ is simple with one extremum at
\beq
A_m=\frac{\Omega\alpha f^2}{M^2+f^2\alpha^2}~.\label{e.Amax}
\eeq
For $f\neq0$ this is a minimum of $V$.

For constant $A$, the potential in $f\geq0$ has three extrema, one maximum at $f=0$ and a minimum and maximum at
\beq
f^2_\pm=\frac13\left(2\pm\sqrt{1+3\kappa^2-3\alpha A(2\Omega-\alpha A)} \right) .\label{e.fpm}
\eeq
The existence of real $f_\pm$ are necessary for a localized soliton solution, which then requires the term under the square root to be non-negative; this provides the following constraint on the amplitude of the gauge field:
\begin{align}
\alpha A \leq \Omega - \sqrt{\Omega_0^2 -1/3}\,,
\label{e.Arvinds_instability}
\end{align}
which is only relevant when $\Omega_0 \geq 1/\sqrt{3}$.

Recall from Eq.~\eqref{e.EnoU} how $f'$ and $A'$ affect the energy. The $f$ profile behaves according to our usual single-particle intuition, but $A$ behaves as if kinetic energy has the opposite sign. Consequently, as the extremum in Eq.~\eqref{e.Amax} is a maximum in $-V$, the dynamics of the system drive $A$ down the $-V$ hill away from the minimum. This implies that for Q-ball solutions, where $f$ begins at a value near one, that $A$ must take values below $A_m$ and is driven by the dynamics to even smaller values. In particular, $2\Omega-\alpha A>0$, so as the system evolves the term under the square-root in Eq.~\eqref{e.fpm} is reduced and the peak in $f$ grows. When $f$ transitions to small values $A_m$ follows suit, which slows the motion of the particle as it approaches $(f=0,A=0)$.

\begin{figure}[t]
\includegraphics[width=0.49\textwidth]{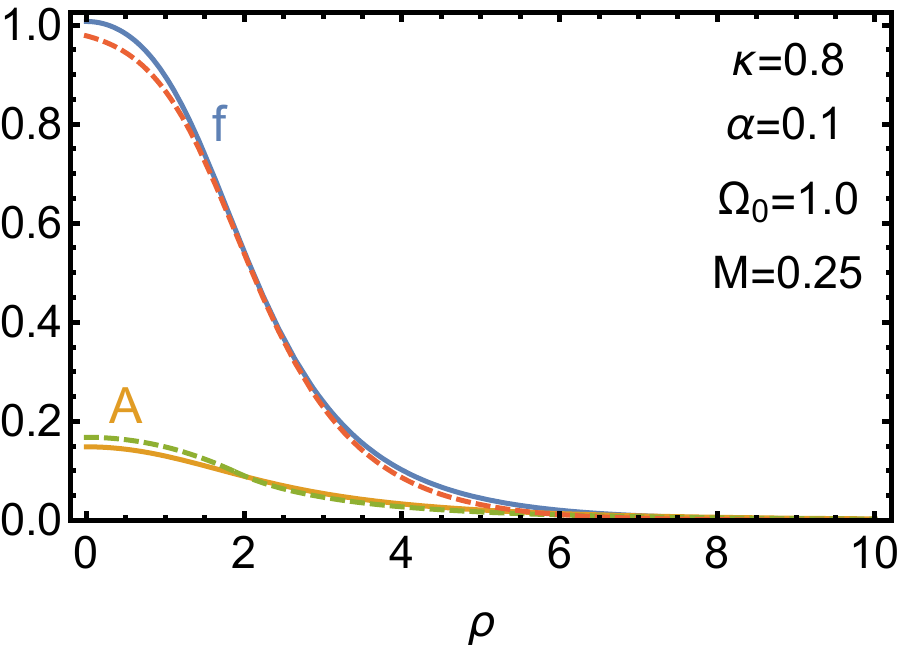}
\includegraphics[width=0.49\textwidth]{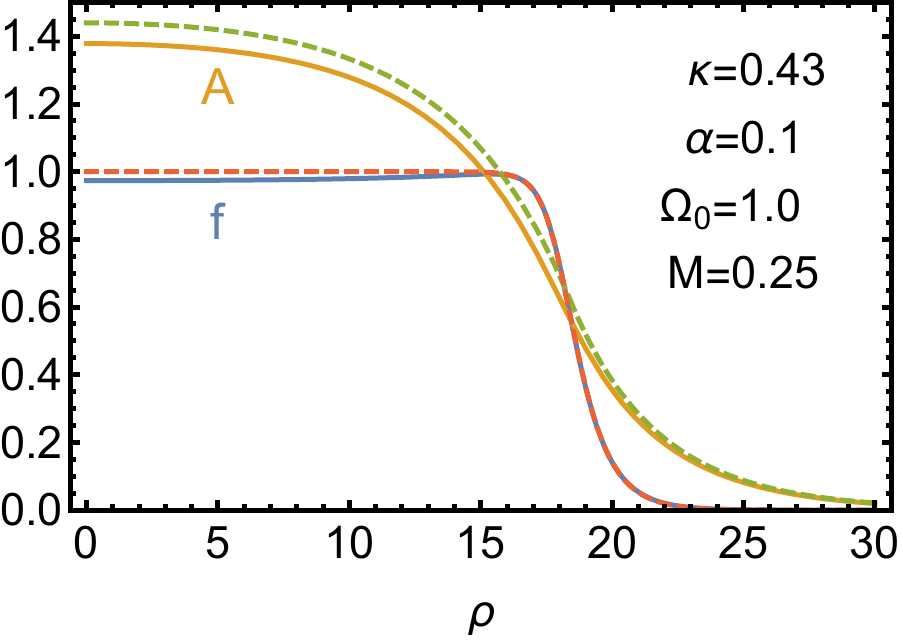}\\
\includegraphics[width=0.49\textwidth]{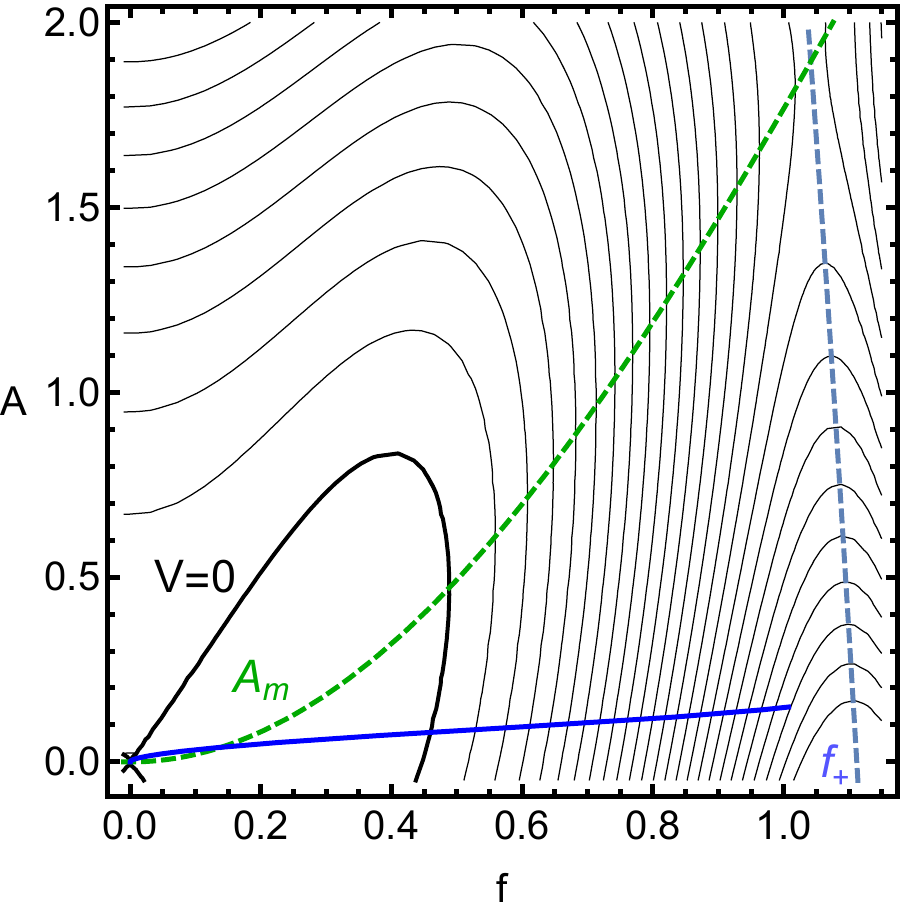}
\includegraphics[width=0.49\textwidth]{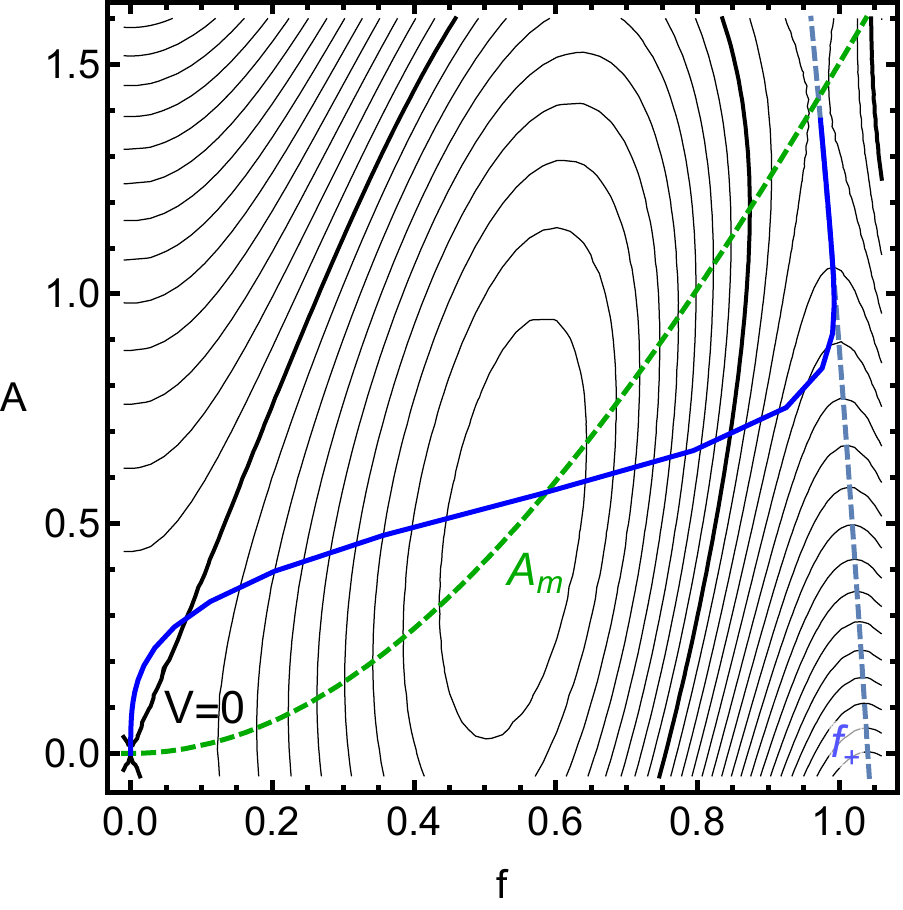}\\
\caption{
Scalar and gauge field profiles for Proca Q-balls (top) and the corresponding trajectories in the potential $V$ (bottom).
Shown are the numerical $f$ profile (blue), the numerical $A$ profile (orange), and the analytic approximations from Eqs.~\eqref{e.thinA} and~\eqref{eq:transition_profile} of the same radius (dashed). A thick (thin) wall Q-ball is shown on the left (right). The values of the $f$ maximum $f_+$ (blue) and $A$ minimum $A_m$ (green) are shown with dashed lines.
}
\label{fig:ball_profiles}
\end{figure}

Two examples of Q-ball trajectories are given in Fig.~\ref{fig:ball_profiles}. On the left we see the scalar and gauge field profiles for a thick-wall Q-ball along with the corresponding trajectory (shown as the thick blue curve) in $V$. A thin-wall profile is shown on the right. In both cases the locations of $f_+$ and $A_m$ within the potential $V$ are shown with dotted blue and green lines, respectively. 

We see that for the thick wall, the trajectory begins at $f\approx1$, a little downhill from $f_+$ causing the particle to roll to smaller $f$. The value of $A$ is less than $A_m$, so the particle rolls uphill toward smaller $A$. When $f$ and $A$ become small enough the trajectory crosses $A_m$ and the gauge field mass begins to dominate the $A$ evolution. 

The thin-wall trajectory begins near the intersection of $f_+$ and $A_m$. This is a point of unstable equilibrium and plays a role similar to a particle resting exactly at $f_+$ in the global Q-ball case. By beginning the evolution close and closer to the equilibrium point the radius of the thin-wall Q-balls can increase without bound. Once the particle begins rolling, being pushed uphill in $V$ to smaller values of $A$, the trajectory remains very close to $f_+$ until the rapid transition in $f$. The particle is slowed by both being above $A_m$ and the usual $f$ direction dynamics, rolling through a valley and then back up to a peak.

\begin{figure}[t]
\includegraphics[width=0.47\textwidth]{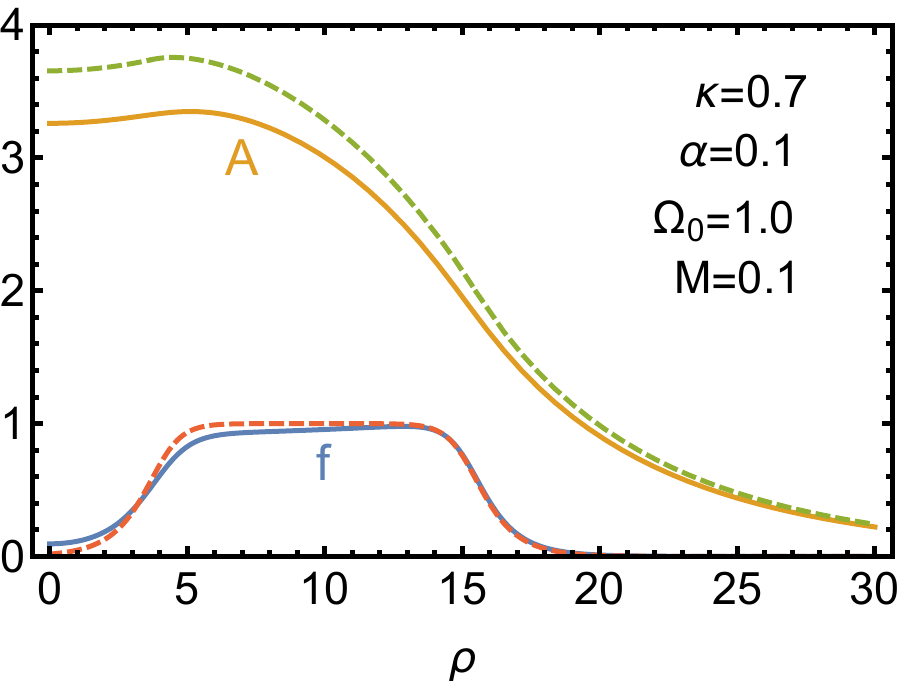}
\includegraphics[width=0.49\textwidth]{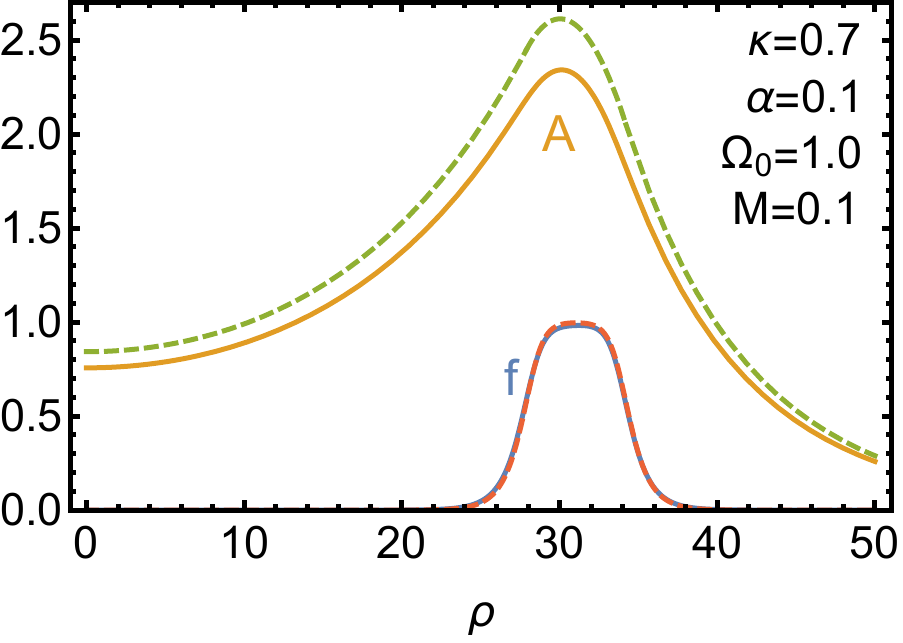}\\
\includegraphics[width=0.49\textwidth]{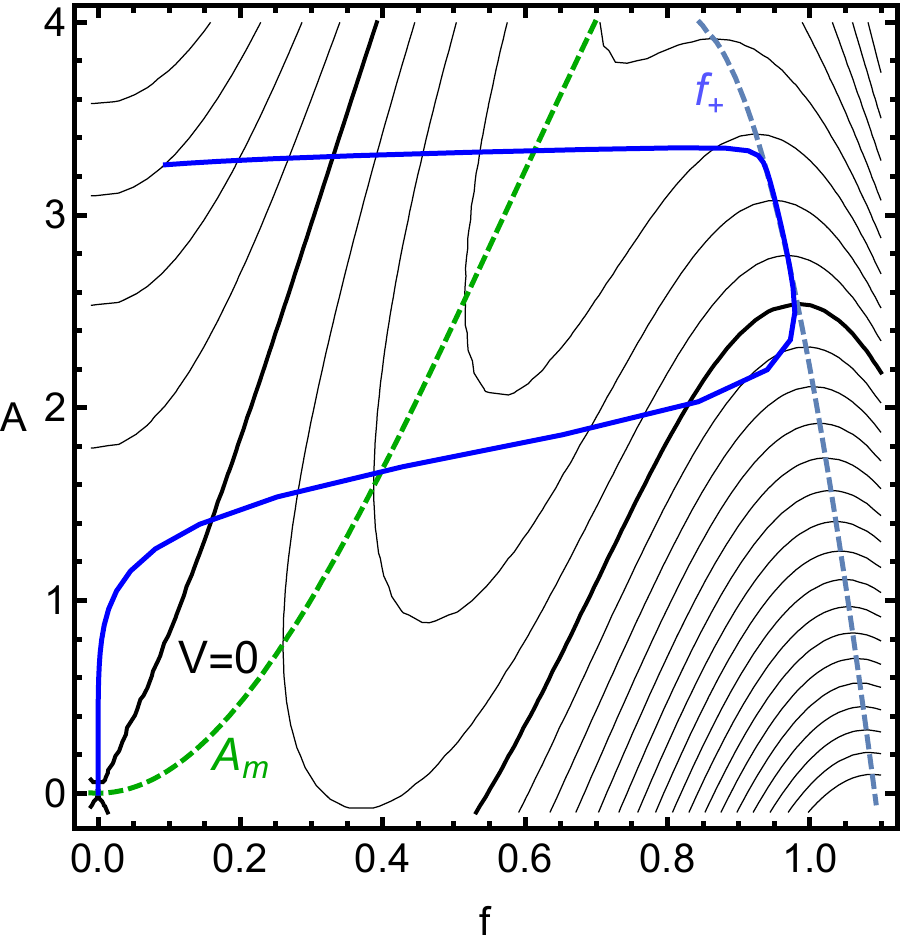}
\includegraphics[width=0.49\textwidth]{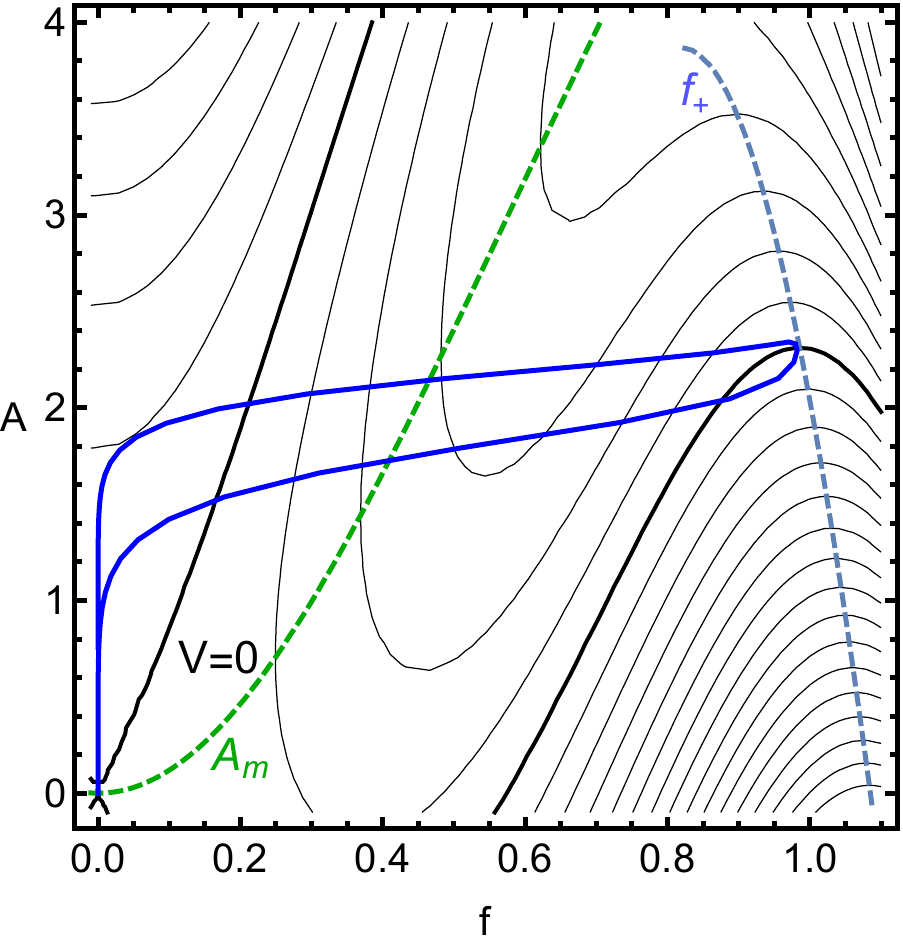}\\
\caption{
Scalar and gauge field profiles for Proca Q-shells. 
Shown are the numerical $f$ profile (blue), the numerical $A$ profile (orange), as well as the analytic approximations from Eqs.~\eqref{eq:Ashell} and~\eqref{eq:fshell} of the same radii (dashed). A wide (narrow) Q-shell is shown on the left (right). The values of the $f$ maximum $f_+$ (blue) and $A$ minimum $A_m$ (green) are shown with dashed lines.
}
\label{fig:shell_profiles}
\end{figure}

For gauged solitons, those with $M=0$, the $A$ profiles must be monotonic~\cite{Lee:1988ag,Heeck:2021zvk}, but this need not hold for $M\neq 0$. This is seen clearly for narrow Proca Q-shells shown on the right side of Fig.~\ref{fig:shell_profiles}. In this case $f$ begins near zero and so the evolution of the system is dominated by the gauge field mass. In other words, with $A>A_m$ the value of the gauge field increases until $f$ transitions to near $f_+$. This change in $f$ suddenly puts $A<A_m$ and so the gauge field is pushed to smaller values. As the particle transitions back toward $f=0$ and $A>A_m$, it is slowed until it comes to rest at the origin of the potential. 

Similar, though milder, behavior is seen for the wide Q-shell on the left side of the figure. The gauge potential starts at a larger value, and increases until the particle rolls in the $f$ direction up to $f_+$. The wide Q-shell trajectory then lies along $f_+$ as $A$ decreases, in contrast to the narrow Q-shell which approaches it only briefly. 

\subsection{Numerical Methods}

Solving the coupled differential equations~\eqref{e.feq} and~\eqref{e.Aeq} is impossible analytically and difficult even numerically. The shooting method discussed by Coleman~\cite{Coleman:1985ki} is quite successful for global Q-balls, but when a gauge coupling and a gauge boson mass are added this approach is tedious at best. To avoid this, we simply solve the boundary value problem as in Refs.~\cite{Heeck:2020bau,Heeck:2021zvk,Heeck:2021gam}.
In order to enforce boundary conditions at $\rho=\infty$ we use a compactified coordinate
\beq
y=\frac{\rho}{1+\rho/a}~,
\eeq
where $a$ is a positive constant. Clearly, $y$ takes values $y\in[0,a]$ and so we can simply require the conditions $f(a)=0$ and $A(a)=0$. The derivatives become
\beq
\frac{\dd\,}{\dd\rho}=\frac{\dd y}{\dd\rho}\frac{\dd\,}{\dd y}=\left(1-\frac{y}{a} \right)^2\frac{\dd \,}{\dd y}~,
\eeq
so the boundary conditions at $y=0$ are still $f'(0)=0$ and $A'(0)=0$ where primes denote a derivative with respect to $y$.
The set of equations
\begin{align}
&\left(1-\frac{y}{a} \right)^4\left(f''+\frac{2}{y}f' \right)+f\left(\kappa^2+\alpha A(\alpha A-2\Omega)-1+4f^2-3f^4 \right)=0\\
&\left(1-\frac{y}{a} \right)^4\left(A''+\frac{2}{y}A' \right)-\alpha f^2\left(\alpha A-\Omega \right)-AM^2=0
\end{align}
can then be solved using finite-element methods.
The success of this method relies heavily on the initial seed function, which should be as close as possible to the exact solution. The field profiles derived below are perfectly suited for this task.

\section{Q-Balls}
\label{sec:balls}

We start our discussion of Proca solitons with Q-balls, a type of solution familiar from both the global~\cite{Heeck:2020bau} ($\alpha\to0$) and gauged~\cite{Heeck:2021zvk} ($M\to0$) special cases. In both cases the scalar profile for large Q-balls is approximately a step function, $f\sim 1 - \Theta (\rho - R^*)$~\cite{Coleman:1985ki,Lee:1988ag}, defined exclusively by the radius $R^*$. Since the gauge boson mass $M$ interpolates, to some degree, between the global and gauged case, it is not surprising that the $f$ profile for large Proca Q-balls also has this shape.
Using this ansatz for $f$ in the $A$ differential equation, Eq.~\eqref{e.Aeq}, yields the solution
\beq
A(\rho)=\frac{\alpha\Omega }{\mu^{2}}\left\{\begin{array}{cc}
\displaystyle 1-\frac{1}{\rho}\frac{(1+MR^\ast)\sinh\left(\rho\mu\right)}{\mu\cosh\left(R^\ast \mu\right)+M\sinh\left(R^\ast\mu\right)}\,, & \rho<R^\ast\\[0.4cm]
\displaystyle\frac{R^\ast\mu-\tanh\left(R^\ast \mu\right)}{\mu+M\tanh\left(R^\ast\mu \right)}\frac{e^{(R^\ast-\rho)M}}{\rho}\,, & \rho\geq R^\ast\\
\end{array}\right. .\label{e.thinA}
\eeq
Here we have required that $A$ and $A'$ be continuous at $\rho = R^*$ and defined the quantity $\mu\equiv\sqrt{\alpha^2+M^2}$, which is the effective mass of the gauge field inside the Q-ball.
 Remarkably, this form for $A$ is a good approximation of the exact solution even beyond the thin-wall or large-radius regime; see for example the green dashed lines in Fig.~\ref{fig:ball_profiles}. 
 The scalar profile can be improved markedly by replacing the step function by the transition profile 
 \begin{align}
 f = \frac{1}{\sqrt{1+2 e^{2 (\rho - R^*)}}}\,,
 \label{eq:transition_profile}
 \end{align}
where the Q-ball radius is defined by $f''(\rho=R^\ast)=0$. This functional form was derived in Ref.~\cite{Heeck:2020bau} as the asymptotic solution for global Q-balls in the limit $R^*\to \infty$. This transition profile remains a good description of Q-ball profiles even for small radii and even in the presence of gauge or Proca fields; see the red dashed lines in Fig.~\ref{fig:ball_profiles}.

These simple profiles for $f$ and $A$ also allow us to calculate analytic expressions for the Q-ball charge and energy; these expressions are not particularly illuminating and are not shown here, but they are compared to the numerical results for several parameter benchmarks below.
Before we can make and compare these predictions, we must determine the radius $R^\ast$ as a function of the potential parameters. This can be achieved via the mapping method of Ref.~\cite{Heeck:2021zvk}.

\subsection{Mapping}

Because the thin-wall solution for $A(\rho)$ from Eq.~\eqref{e.thinA} is a good approximation of most exact gauged Q-ball solutions---as shown below---we can use it to estimate how much $\alpha A$ changes when $f$ transitions from $1$ to $0$. We find
\beq
|\alpha A'(R^\ast)|=\alpha^2\frac{\Omega(1+MR^\ast)}{\mu^2R^{\ast2}}\left|\frac{\tanh(\mu R^\ast)-\mu R^\ast}{\mu+M\tanh(\mu R^\ast)}\right| < \frac{\alpha\Omega}{3}~,
\eeq
which implies that $\alpha A'\ll1$ for small $\alpha\Omega$ during the $f$ transition from interior to exterior, so $A$ is approximately constant.
In fact, we show below that this mapping is qualitatively, and often quantitatively, accurate even beyond the small $\alpha\Omega$ limit.
 In the region around $\rho \sim R^*$, the $f$ equation in Eq.~\eqref{e.feq} takes the form
\beq
(\rho^2f')'=\frac{\rho^2}{\Phi_0^2(m_\phi^2-\omega_0^2)^2}\frac{\dd U}{\dd f}-\rho^2\left[\Omega-\alpha A(R^\ast)\right]^2f
\eeq
which is exactly the form of the potential for the \emph{global} Q-ball equation with the global value of $\Omega=\Omega_G$ given by
\beq
\Omega_G=\Omega-\alpha A(R^\ast)\,.
\eeq
This makes clear that the $f$ transition profiles for Proca Q-balls can be identified with particular transition profiles for \emph{global} Q-balls; for large $R^\ast$, these are simply the transition profiles given in Eq.~\eqref{eq:transition_profile}. These profiles only match when the amount of friction is the same for both the global and gauged cases, which implies that the radius $R^\ast$ is the same for each. Therefore, if the $R^\ast$ dependence of the global Q-ball parameter $\Omega_G(R^\ast)$ is known, we can determine the $R^\ast$ dependence of the Proca Q-ball $\Omega(R^\ast)$ via
\begin{align}
\boxed{
\Omega(R^\ast)= \Omega_G(R^\ast)\left[1-\frac{\alpha^2}{\mu^{2}R^\ast}\frac{R^\ast\mu-\tanh(R^\ast\mu)}{\mu+M\tanh(R^\ast\mu)} \right]^{-1} ,}\label{e.gaugeOm}
\end{align}
where we have used the thin-wall formula given in Eq.~\eqref{e.thinA} for $A(R^\ast)$.
As quick sanity checks we can easily verify that $\Omega(R^\ast)\to  \Omega_G(R^\ast)$ for either $\alpha \to 0$ or $M\to \infty$ as it should.
In the limit $M\to 0$, the success of this mapping relation has been demonstrated in Ref.~\cite{Heeck:2021zvk}.

Equation~\eqref{e.gaugeOm} provides a powerful mapping from global Q-balls---for which the relation $\Omega_G(R^\ast)$ between frequency and radius is much easier to obtain both analytically and numerically---and Proca Q-balls.
This rather simple result provides amazingly accurate analytic descriptions using the global $\Omega_G (R^*)$ relation for our sextic potential found in Ref.~\cite{Heeck:2020bau}.

The solution of the differential equations is approximated as follows: Eq.~\eqref{e.gaugeOm} provides the radius of the Proca Q-ball given the known relationship $\Omega_G (R^\ast)$ from the global Q-ball (Ref.~\cite{Heeck:2020bau}). 
The scalar profile $f(\rho)$ is taken to be the transition profile of global Q-balls of Eq.~\eqref{eq:transition_profile}; this is well motivated around $\rho\sim R^\ast$ for large $R^\ast$ but happens to be a very good approximation for all other cases as well. Finally, the gauge profile $A(\rho)$ is taken from Eq.~\eqref{e.thinA}.
Below we compare these approximations with exact numerical results and show that they are remarkably good, even far beyond their expected region of validity.

\subsection{Discussion and Comparison}

In Fig.~\ref{fig:ball_profiles} we show two examples of predicted profiles for $f$ and $A$ together with the exact numerical results. The agreement is particularly good for small $\alpha$ and large $R^*$, the approximations that lead us to our mapping formula and profile functions, but are still useful for large $\alpha$ and small $R^*$.

What then are the implications of the mapping relation in Eq.~\eqref{e.gaugeOm}?
In general, the predicted relation between the frequency $\Omega$ and the Q-ball radius $R^*$ is far more complex than for either global or gauged Q-balls. However, in the limit of large $R^*$, where our analytic approximations are best, the behavior is quite simple. Using the large-$R^*$ global Q-ball relation $\Omega_G^2 \simeq \Omega_0^2 + 1/R^*$~\cite{Heeck:2020bau} in Eq.~\eqref{e.gaugeOm}, we find 
\begin{align}
\kappa (R^*\to \infty) \to \frac{\alpha \Omega_0}{M} 
\label{eq:large_R_kappa}
\end{align}
for $M>0$.
Since stable localized Q-balls require $\kappa \leq 1$ (corresponding to $\omega\leq m_\phi$), we immediately conclude that Proca Q-balls with $M < \alpha \Omega_0$ exhibit a maximal radius just like gauged Q-balls~\cite{Heeck:2021zvk}, whereas Proca Q-balls with $M > \alpha \Omega_0$ can potentially be arbitrarily large just like global Q-balls~\cite{Heeck:2020bau}.\footnote{Additional relations need to be satisfied to evade a maximal radius, in particular Eq.~\eqref{e.Arvinds_instability}.} This quantifies the statements in the introduction that heavy (light) gauge bosons give rise to solitons that resemble global (gauged) ones.

\begin{figure}
\begin{center}
\includegraphics[width=0.47\textwidth]{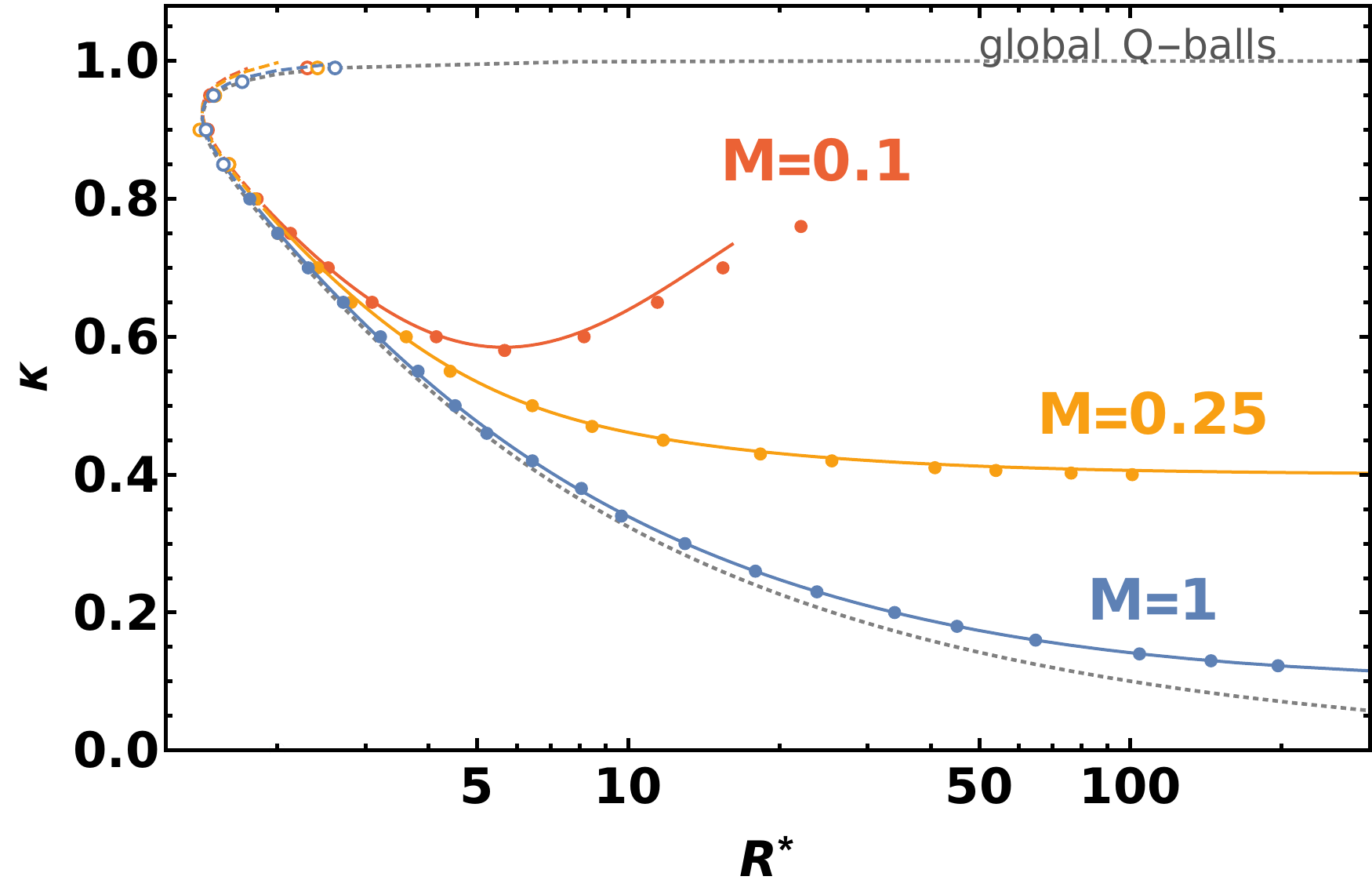}
\includegraphics[width=0.47\textwidth]{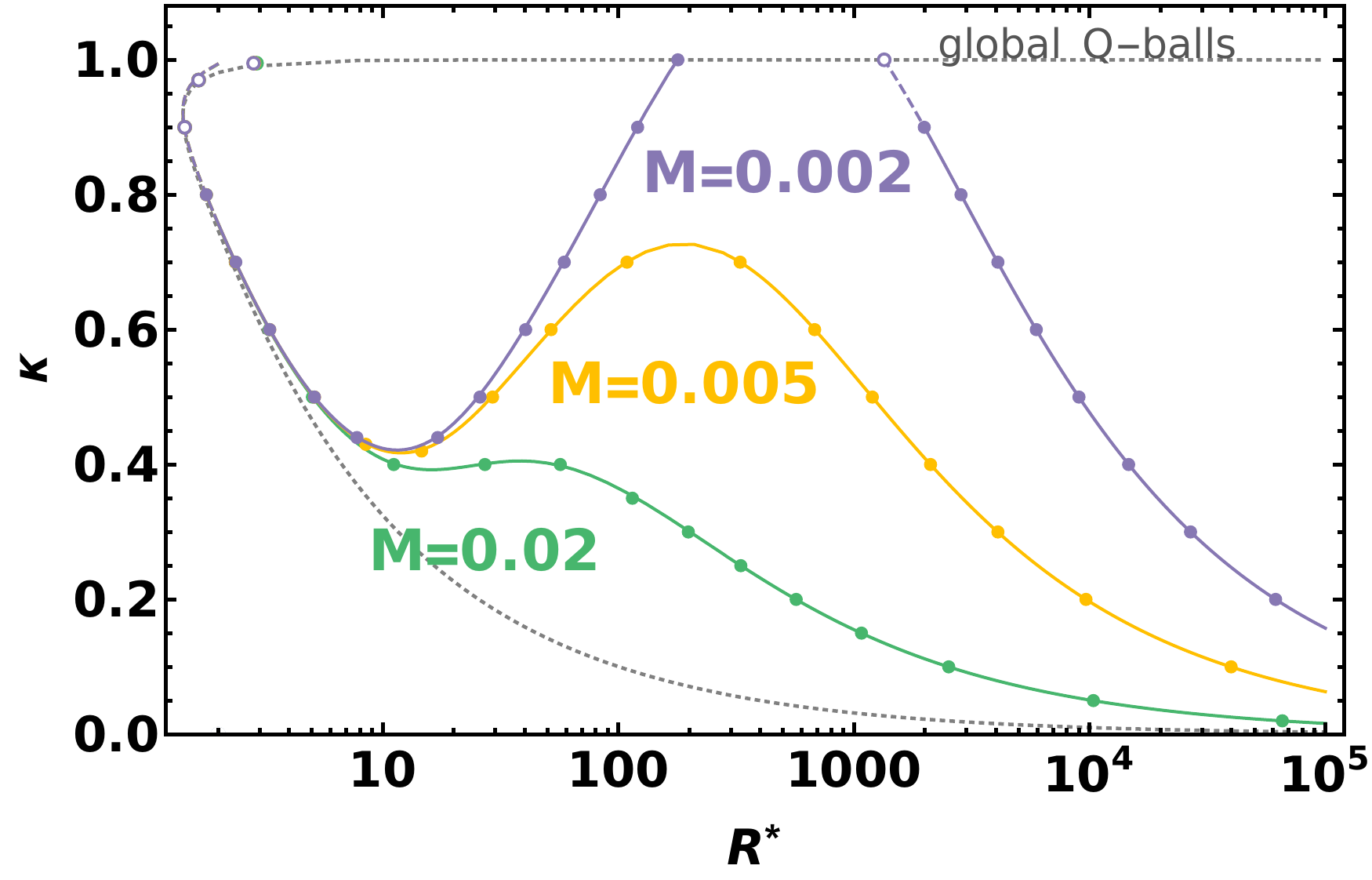}
\includegraphics[width=0.47\textwidth]{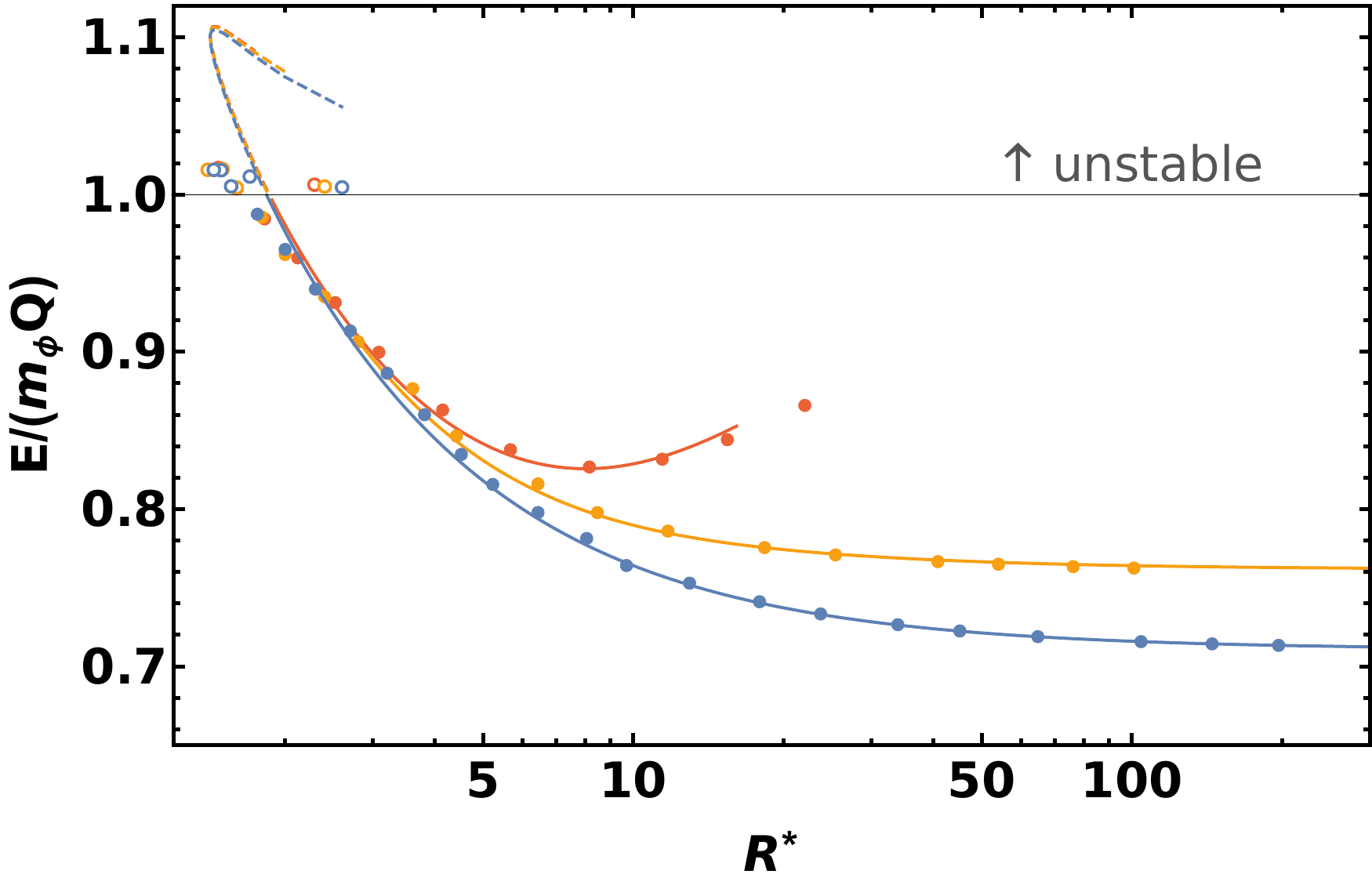}
\includegraphics[width=0.47\textwidth]{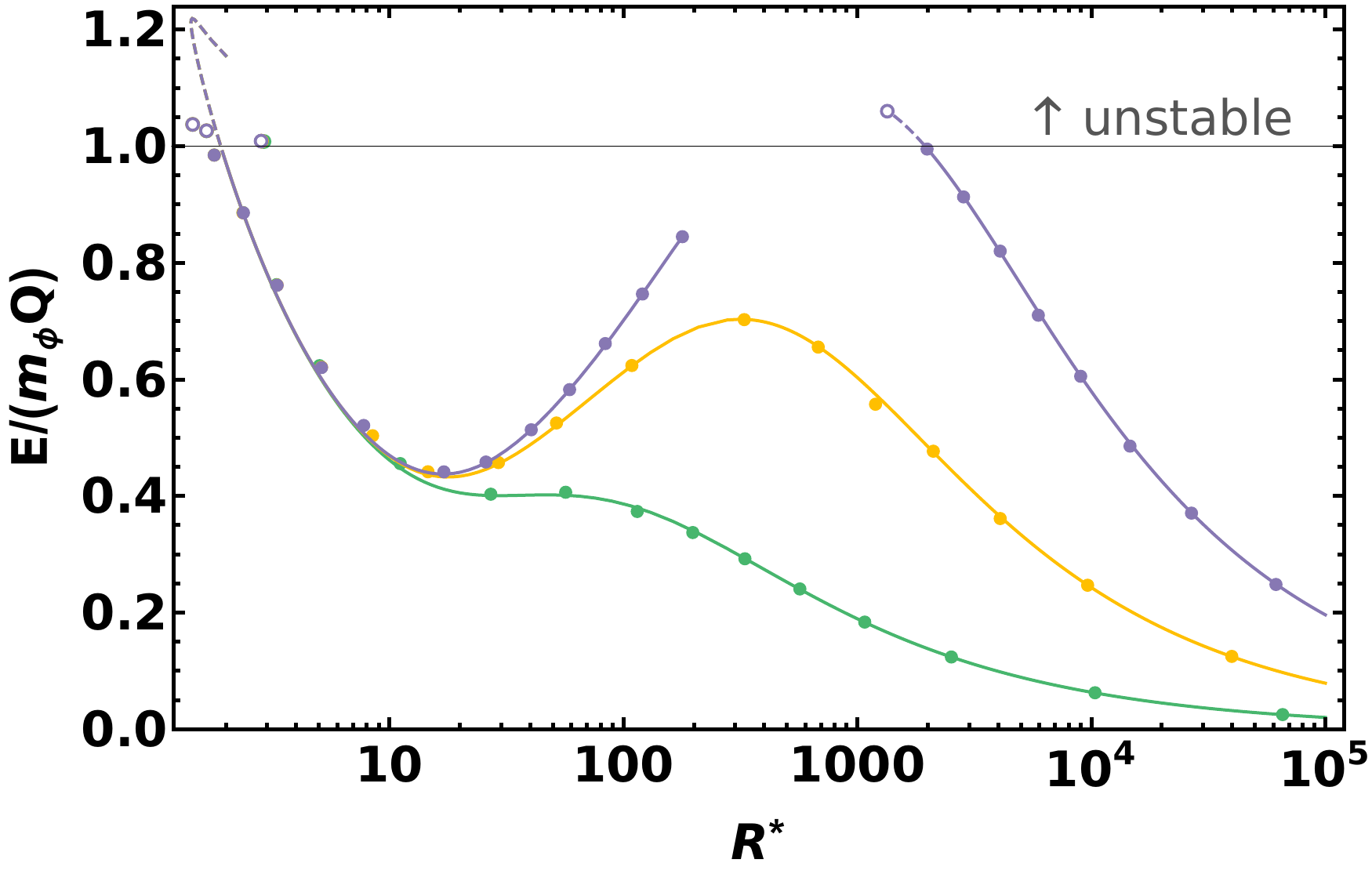}
\includegraphics[width=0.48\textwidth]{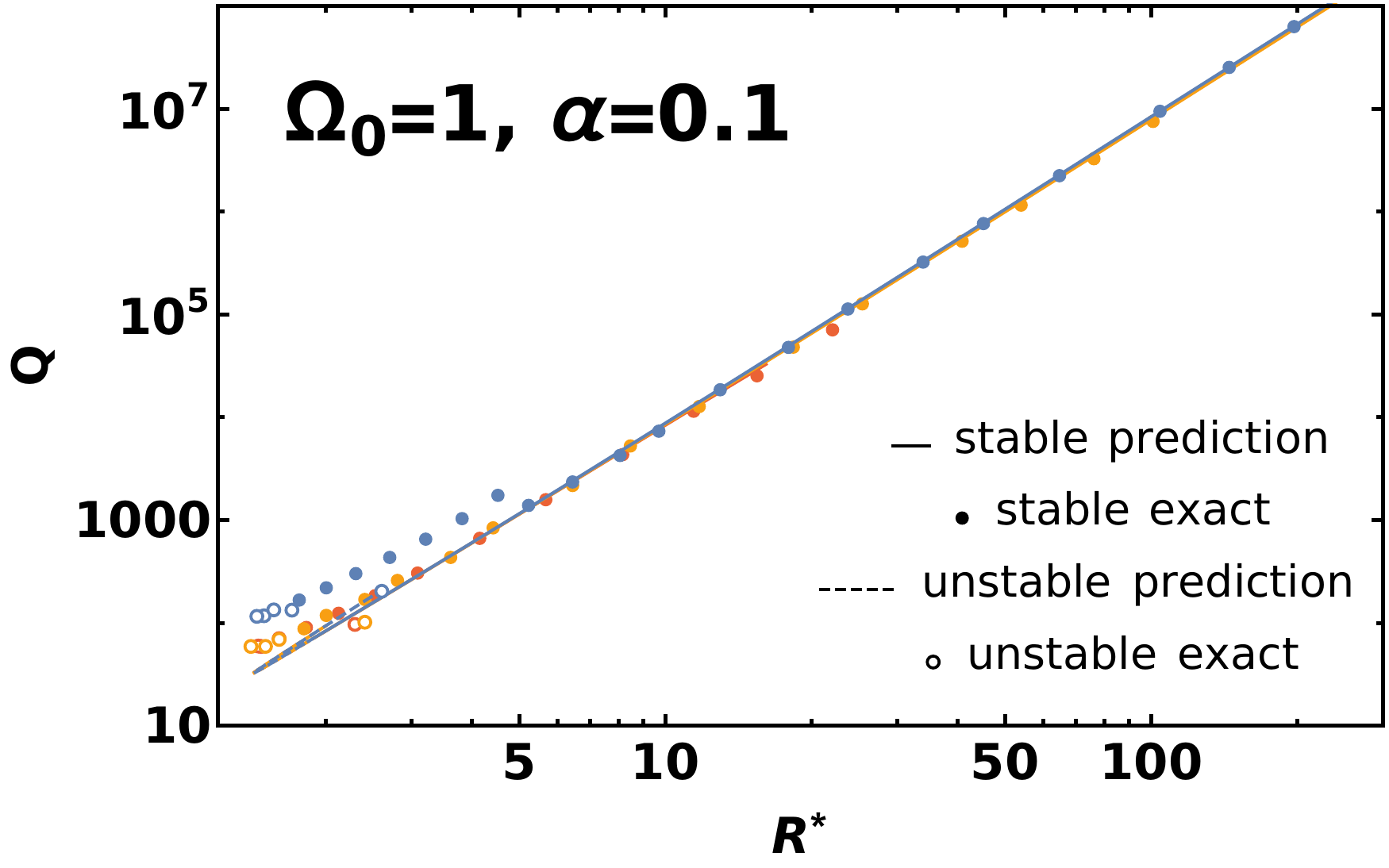}
\includegraphics[width=0.48\textwidth]{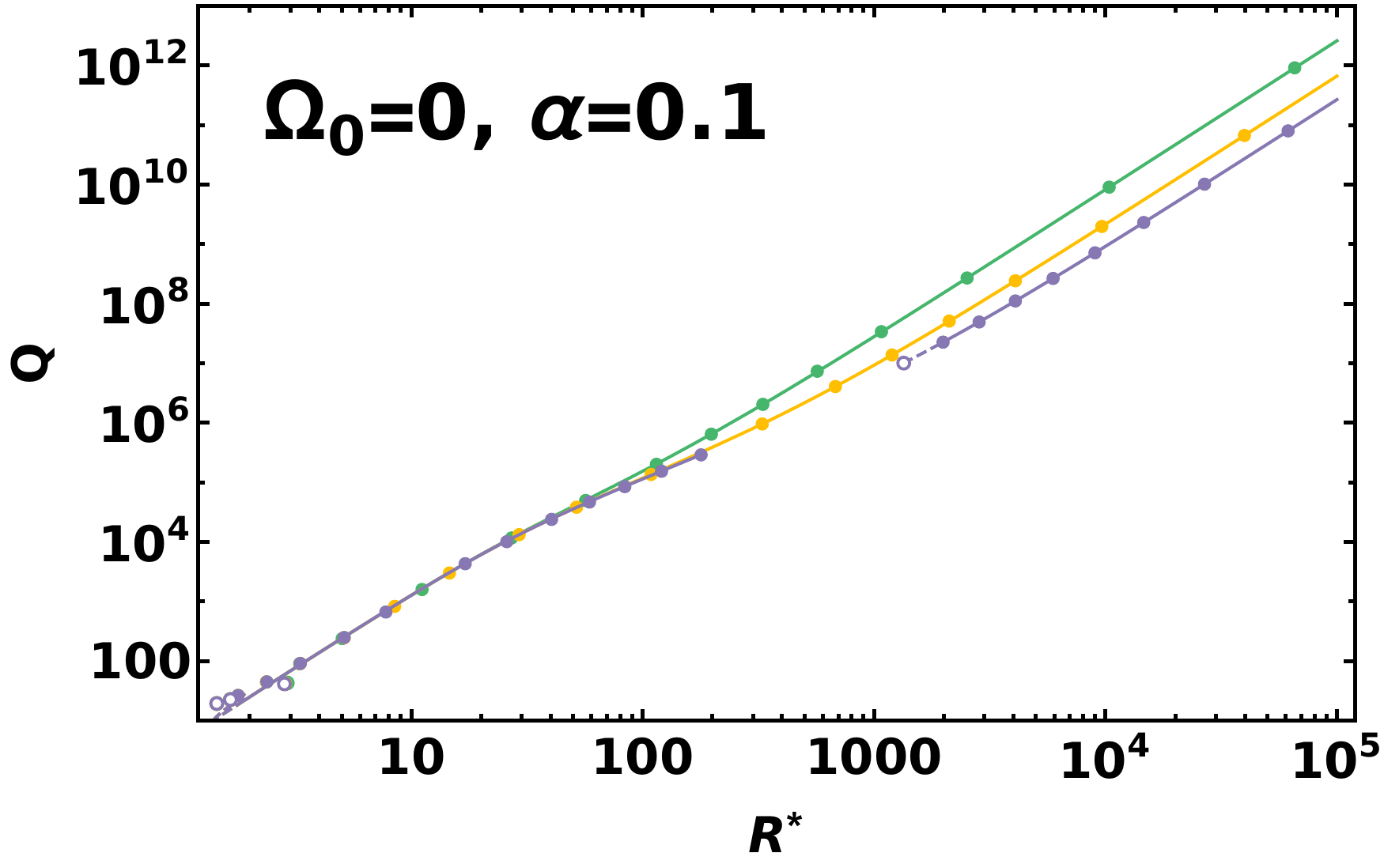}
\end{center}
\caption{
Proca Q-ball solutions vs.~$R^\ast$ for $\Omega_0=1$ (left column), $\Omega_0=0$ (right column), and various $M$.
Top row: $\kappa (R^\ast)$ for Proca Q-balls. The solid lines correspond to our predictions using the mapping of Eq.~\eqref{e.gaugeOm}, the dots to exact numerical solutions, and the gray dotted line to the global Q-ball case shown for comparison.
Middle row: $E/(m_\phi Q)$ vs.~$R^\ast$ using the same benchmark parameters as above. Q-balls are unstable for $E/(m_\phi Q) > 1$.
Bottom row: $Q$ vs.~$R^\ast$.  The solid lines are our analytic approximations.
}
\label{fig:exact_solutions}
\end{figure}

Some benchmark scenarios for Proca Q-balls are shown in Fig.~\ref{fig:exact_solutions}. In addition to exact numerical solutions (circles), we show the analytical predictions (curves) for $\kappa (R^\ast)$ from the mapping relation in Eq.~\eqref{e.gaugeOm} --- using the full numerical relation for the global $\kappa (R^\ast)$ from  Ref.~\cite{Heeck:2020bau} --- as well as the predictions for $E$ and $Q$.
The $\kappa (R^\ast\to \infty)$ behavior from Eq.~\eqref{eq:large_R_kappa} is clearly shown for the $M=0.25$ and $M=1$ benchmarks in the left column of the figure.
The benchmark point $M=0.1$ exhibits an instability at $R^\ast \sim 23$ due to violation of Eq.~\eqref{e.Arvinds_instability} and hence has a maximal radius.\footnote{Using Eq.~\eqref{e.Arvinds_instability} together with $A = A(0)$ from Eq.~\eqref{e.thinA} and the mapping relation predicts this instability at $R^\ast > 16$, in qualitative agreement with the numerical result.} This parameter point also supports Q-shells, which are discussed in Sec.~\ref{sec:shells}.

The behavior of $\kappa (R^\ast)$ can be more complex than the left column of Fig.~\ref{fig:exact_solutions} suggests; for small $\Omega_0$, $\kappa (R^\ast)$ can have a local maximum at large $R^\ast$ and thus up to three branches of Q-ball solutions for a given $\kappa$. This is illustrated by the benchmarks in the right column. Because $\Omega_0=0$, these Proca Q-balls resemble global Q-balls for very large $R$ and have no maximal charge or radius. However, the case $M=1/500$ has an instability region around $R^\ast \in (200,2000)$ defined by $\kappa >1$ and $E>m_\phi Q$. These disconnected stability regions are a novel feature of Proca Q-balls.

These benchmarks illustrate the success of the mapping formula Eq.~\eqref{e.gaugeOm} and the analytic profiles for $f$ and $A$. It is not difficult to derive analytic expressions for special points in parameter space, such as the local minima and maxima of $\kappa (R^\ast)$ or instability regions where $\kappa (R^\ast)> 1$, which are left as an exercise to the reader.
One interesting special case deserves mentioning, though: there exists a region of parameter space where the local minimum $\kappa_\text{min}(R^\ast)$ is larger than $1$; the valid Q-ball radii then do not start at $R^\ast \sim 1$ as in almost all other cases, but rather at $R^\ast \gg 1$. The formation of these extra-large Q-balls, as well as those with disconnected stability regions, will be studied in future work.

\section{Q-Shells}
\label{sec:shells}

\emph{Global} solitons in our sextic potential~\eqref{eq:sextic_potential} only exist in the form of Q-balls~\cite{Heeck:2020bau}; the presence of a gauge field gives rise to a qualitatively different kind of soliton: Q-shells~\cite{Heeck:2021gam}. Similar Q-shells are expected to also exist in the Proca case, at least for small gauge boson masses. While we do find these soliton solutions that match to the pure gauge Q-shells when $M\to0$, we also discover Q-shells with no $M\to0$ analogue. Thus we see that, like in the Q-ball case, the introduction of a Proca mass term leads to new classes of soliton solutions. 

Following Ref.~\cite{Heeck:2021gam} we can approximate the scalar profile of large-radii Q-shells by
\begin{align}
f(\rho) = \begin{cases}
0\,, & \rho < R_< \,,\\
1\,, & R_< \leq \rho \leq R_>\,,\\
0\,, & R_> < \rho\,,
\end{cases}
\end{align}
specified by the two radii of the shell. With this ansatz we can solve the $A$ equation of motion and find
\begin{align}
A(\rho) = \begin{cases}
\displaystyle A_<\frac{R_<}{\rho}\frac{\sinh (M\rho)}{\sinh(MR_<)} \,, & \rho<R_<\,,\\
\displaystyle\frac{\alpha\Omega }{\mu^2}-\frac{A_1}{\rho}\sinh(\mu \rho) -\frac{A_2}{\rho}\cosh(\mu\rho)\,, & R_<\leq\rho\leq R_>\,,\\
\displaystyle A_>\frac{R_>}{\rho}e^{(R_>-\rho)M} \,, & R_><\rho \,,
\end{cases} 
\label{eq:Ashell}
\end{align}
with $A_<$, $A_1$, $A_2$, and $A_>$ determined by demanding continuity of $A$ and $A'$ at the two radii. Similar to the Q-ball case, this profile is a remarkably good description of all Q-shells, even when the radii are small; see Fig.~\ref{fig:shell_profiles}. The $f$ profile can be improved by replacing the step function by a double-transition profile~\cite{Heeck:2021gam}
\begin{align}
f(\rho) = \frac{1}{\sqrt{1+ 2 e^{2(R_<-\rho)}}}\,\frac{1}{\sqrt{1+ 2 e^{2(\rho-R_>)}}} \,.
\label{eq:fshell}
\end{align}
These profiles for $f$ and $A$ give the correct qualitative behavior, as illustrated in Fig.~\ref{fig:shell_profiles}.
It remains to determine the two radii $0 < R_< < R_>$ as a function of the potential parameters.

By requiring continuity in $A$ and $A'$ and $R_<$ and $R_>$ one obtains four equations relating the six parameters $A_<$, $A_>$, $A_1$, $A_2$, $R_<$, and $R_>$. To fully specify the system we need two more independent equations that relate the parameters. Similar to the gauged Q-shell case~\cite{Heeck:2021gam}, we use the energy-due-to-friction relation given in Eq.~\eqref{e.WorkFric} to deduce the following approximate relations:
\begin{align}
R_<&\simeq -\left[ \kappa^2-\alpha A_<\left(2\Omega-\alpha A_<\right)\right]^{-1},\label{e.Rg}\\
R_>&\simeq \left[ \kappa^2-\alpha A_>\left( 2\Omega-\alpha A_>\right)\right]^{-1} .\label{e.Rl}
\end{align}
These equations are sufficient to fully specify the system. Though difficult to solve analytically, by treating them numerically the Q-shell radii can be predicted as a function of the potential parameters. 

While general solutions are challenging, we can extract some information from the limit of very large radii. In this limit we consider both $R_<$ and $R_>$ as becoming infinite, so the results only apply to narrow Q-shells. In this limit of infinite radius we find
\beq
\kappa^2 \sim \frac{\displaystyle \alpha^2 - M^2+\sqrt{(\alpha^2 - M^2)^2 + 8 M \alpha^2 \Omega_0^2}}{2M}\,;\label{e.kapInf}
\eeq
this the value of $\kappa$ at which we expect the narrow Q-shell radii to grow without bound. 

\begin{figure}[t]
\includegraphics[width=0.496\textwidth]{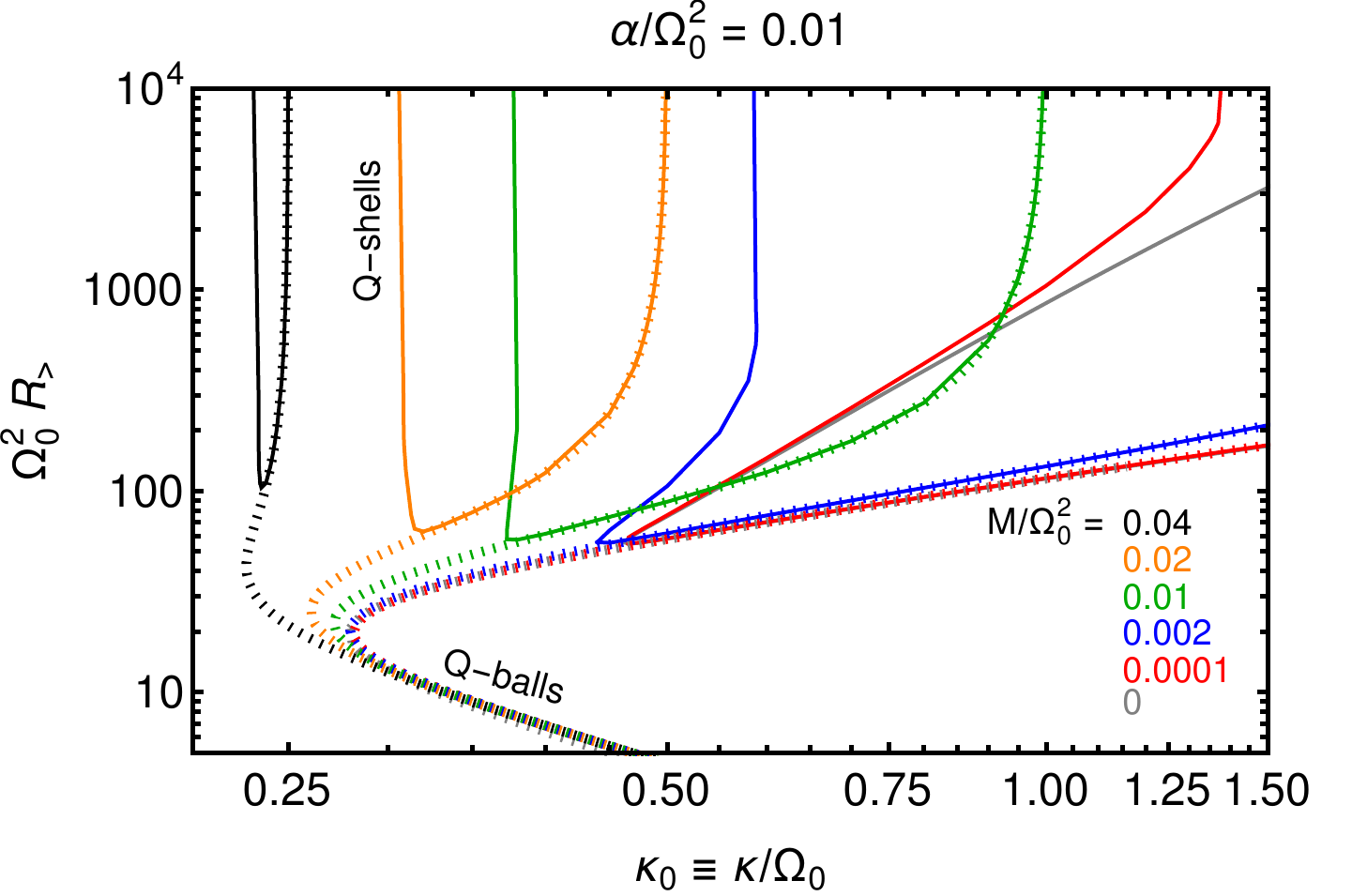}
\includegraphics[width=0.496\textwidth]{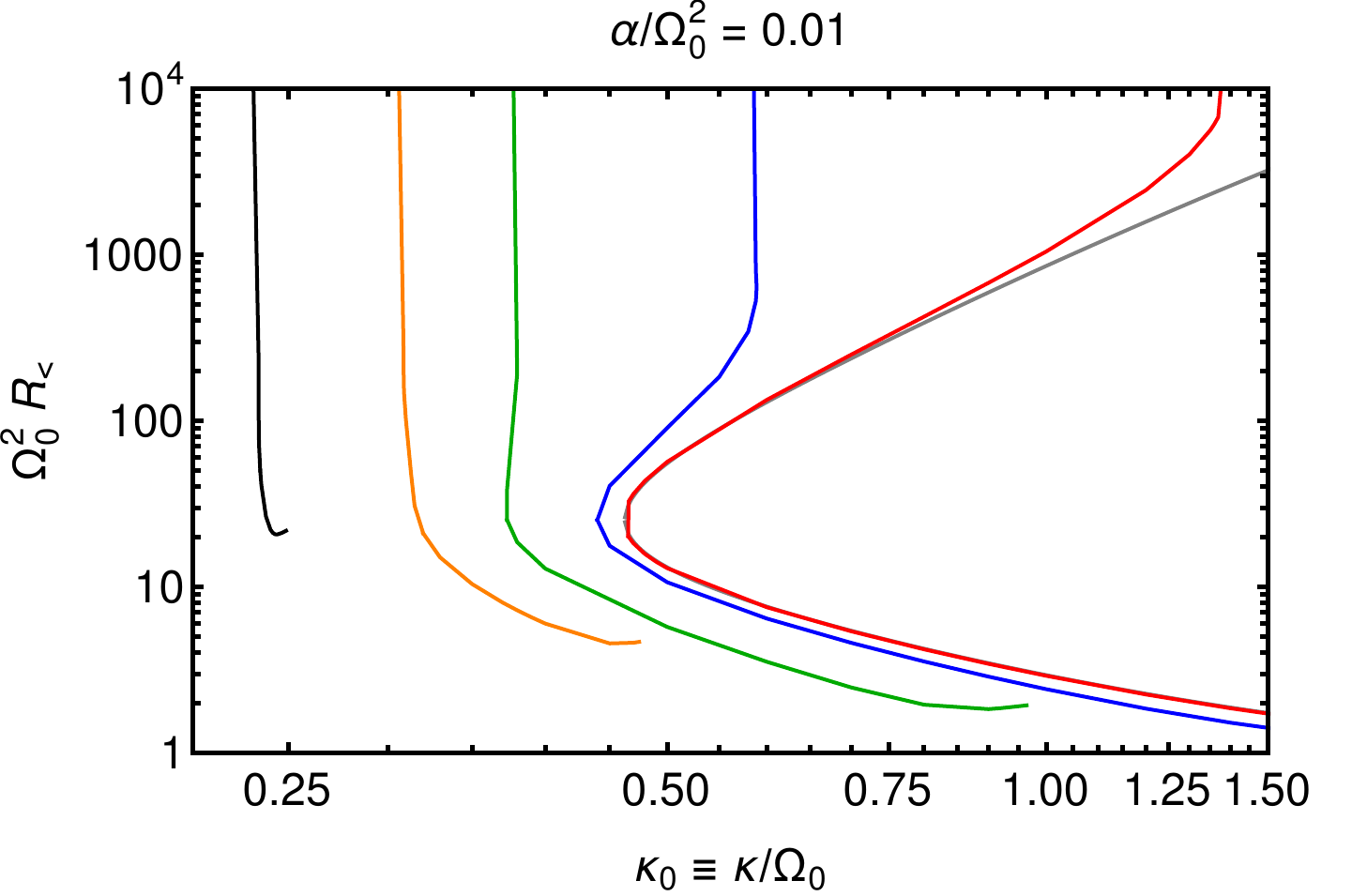}
\caption{
Predictions for the Q-shell radii $R_>$ (left) and $R_<$ (right) based on Eqs.~\eqref{e.Rg} and \eqref{e.Rl} for several values of the gauge-boson Proca mass $M$. The narrow Q-shell radii $R_{>,<}$ diverge at $\kappa$ given by Eq.~\eqref{e.kapInf}; the wide Q-shell radii $R_>$ are degenerate with the Q-ball radii (dashed) and both end at a radius that is determined by Eq.~\eqref{e.Arvinds_instability} for a given $\Omega_0$, not shown in the plot.
}
\label{fig:Qshell_radius_prediction}
\end{figure}

In Fig.~\ref{fig:Qshell_radius_prediction} we display the predictions for outer (left) and inner (right) radii for Q-shells of increasing $M$. The equations in the large-radius limit can be written in terms of rescaled parameters like $\alpha/\Omega_0^2$ such that all explicit dependence on $\Omega_0$ vanishes, just like in the gauged case~\cite{Heeck:2021gam}. The Q-ball predictions are also shown in wide dashes in the left plot. Note that the wide Q-shells continue to follow exactly along the thin-wall Q-balls. We can also see from both plots that the narrow Q-shells' radii diverge at the values of $\kappa$ given by Eq.~\eqref{e.kapInf}. We also note that the narrow Q-shells' curves seem to be bounded, in $\kappa$, from below by the Q-ball $\kappa_\text{min}$. Consequently, when the parameters are such that the Q-ball prediction given by the mapping relation does not fold back in $\kappa$, but has a one-to-one relation between $\kappa$ and $R^\ast$, there are no Q-shell solitons.

\begin{figure}[t]
\includegraphics[width=0.49\textwidth]{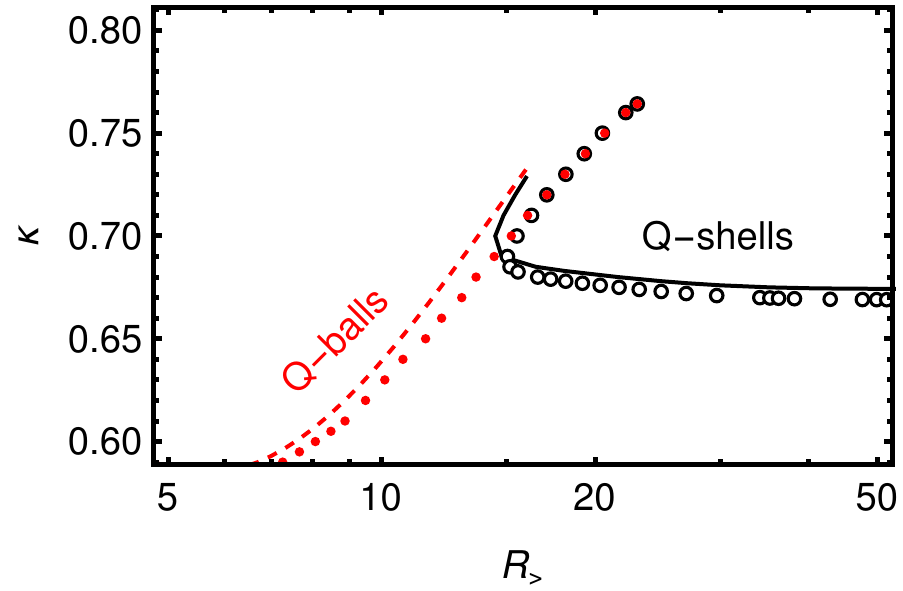}
\includegraphics[width=0.49\textwidth]{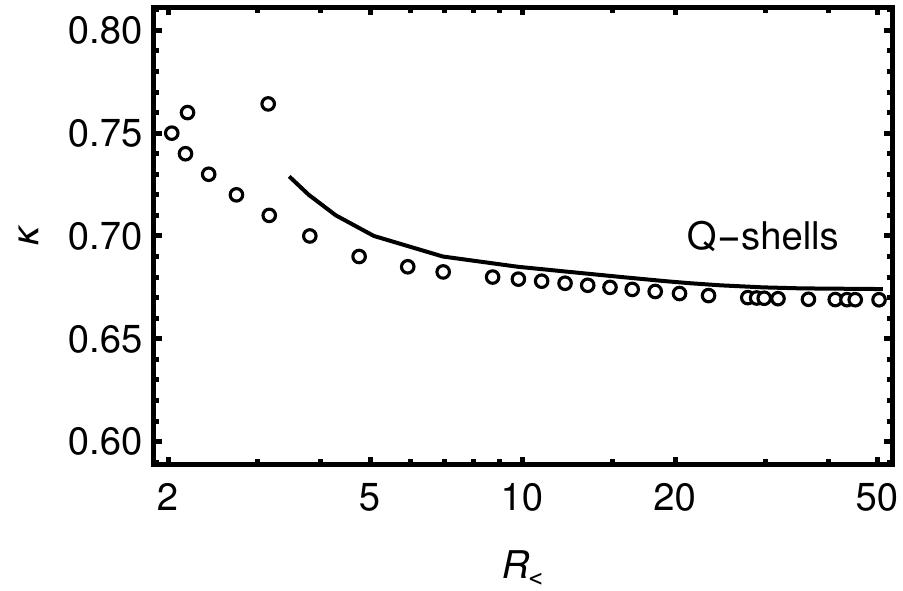}\\
\includegraphics[width=0.49\textwidth]{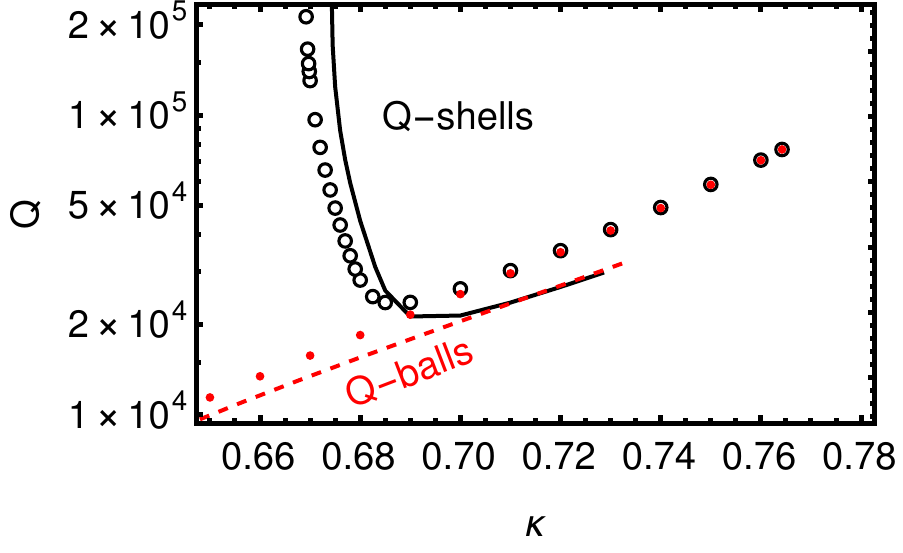}
\includegraphics[width=0.49\textwidth]{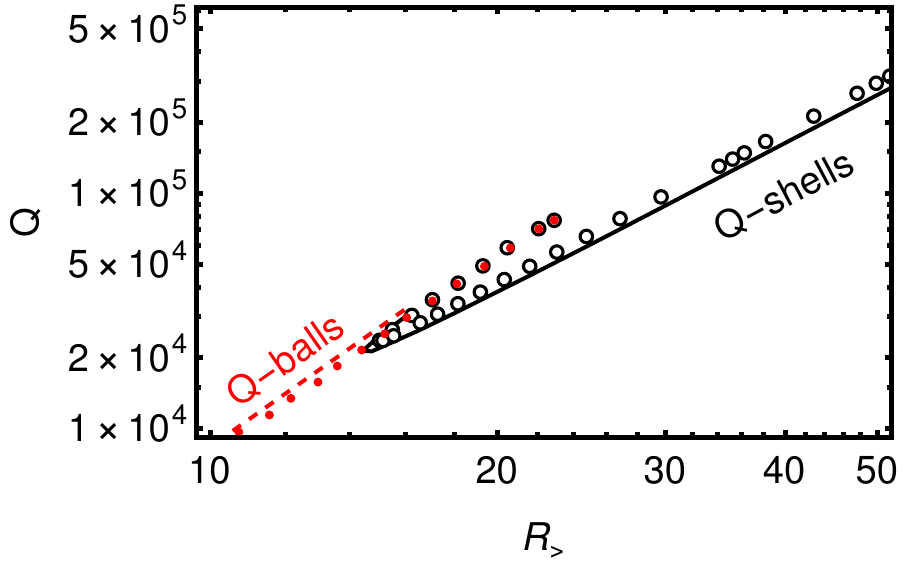}\\
\includegraphics[width=0.49\textwidth]{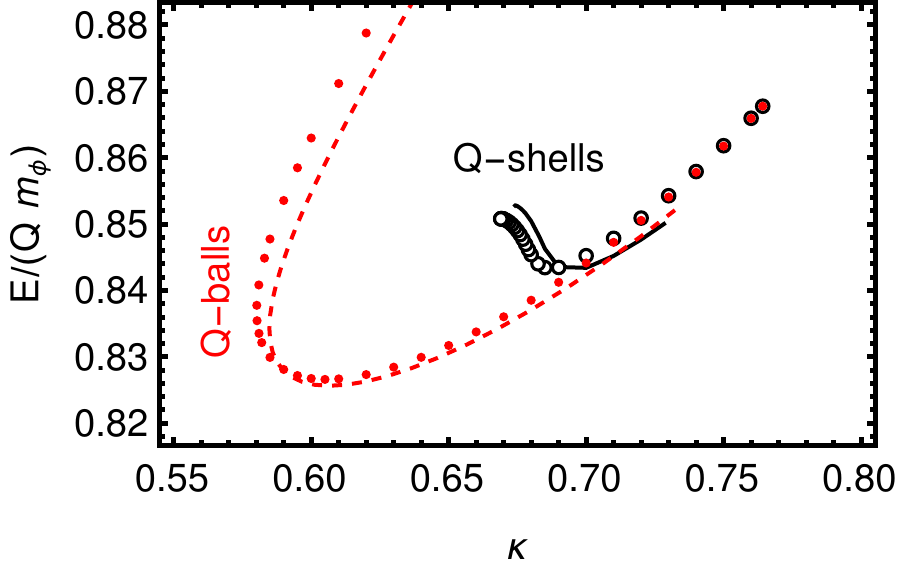}
\includegraphics[width=0.49\textwidth]{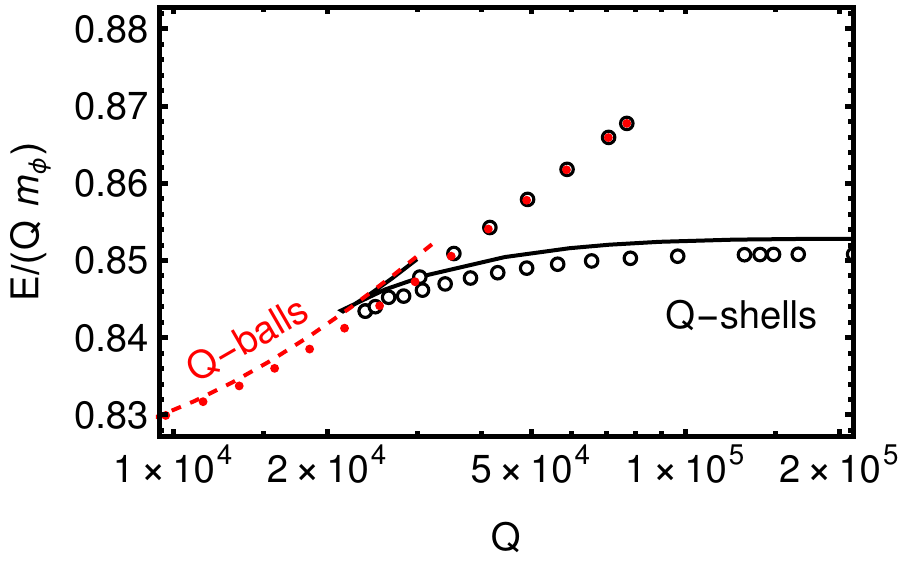}
\caption{
Proca solitons for the benchmark point $\alpha = M =\Omega_0/10 = 1/10$, $\phi_0=m_\phi$. The red points correspond to exact Q-ball solutions, the dashed red line to our prediction from Eqs.~\eqref{e.Rg} and \eqref{e.Rl}; the full Q-ball branch is shown in Fig.~\ref{fig:exact_solutions}. 
The black circles are exact Q-shell solutions and the black line our analytical approximation.
Notice that the wide Q-shells are degenerate with Q-balls and cease to exist for $\kappa > 0.76$ due to the instability described in Eq.~\eqref{e.Arvinds_instability}.
}
\label{fig:bench5shells}
\end{figure}

We compare the predicted Q-shell characteristics with exact numerical results in Fig.~\ref{fig:bench5shells}. This benchmark is for $\alpha=0.1$, for which there are no Q-shells when $M\to0$~\cite{Heeck:2021gam}. This tests our predictions away from the small $M$ limit, where both $M$ and $\alpha$ have the same magnitude. The figure shows impressive agreement between the theoretical predictions and the exact numerical results. We see that the wide Q-shells and thin-wall Q-balls always lie together. This suggests that we can use the numerically simpler Q-ball mapping relation to obtain the wide Q-shell characteristics, other than the inner radius. 

We also note that, as predicted, $E/(m_\phi Q)$ (which can be thought of as measuring stability) is smaller, at fixed $Q$, for narrow Q-shells than for the other solitons. This would seem to imply that narrow Q-shells may be, in this sense, the more stable solitons for a given $Q$. These more stable Q-shells also have a larger outer radius. These results suggest that the cosmological production of Proca solitons may favor Q-shells in some cases, though a thorough analysis is beyond the scope of this work. 

\section{Conclusion\label{s.con}}

Non-topological solitons in complex scalar field theories have long been discussed in both the case of an underlying global $U(1)$ symmetry and a gauged $U(1)$. Here, we present the inaugural study of solitons made up of complex scalars coupled to a massive Proca gauge boson. The resulting solitons are similar to the global ones, i.e.~Q-balls, for large Proca masses and similar to gauged ones, Q-balls and Q-shells, for small Proca masses. For intermediate gauge boson masses the solitons show unique features such as Q-balls with extremely large minimal radii and charge and Q-shells with arbitrarily large radii.
We have provided powerful analytic approximations for the scalar and gauge profiles for these Proca solitons that allow insights into their behavior and also make possible efficient numerical studies.

Open avenues for further study of these novel solitons include their stability with respect to decay into smaller solitons as well as the reverse issue of soliton formation from smaller solitons and individual scalars. The possibility of transitions between Q-balls and Q-shells of equal charge is also intriguing. Addressing these questions is likely to precede a complete understanding of how these solitons might be produced in the early universe and how they might persist as a component of the Universe's dark matter.

\section*{Acknowledgements}
This work was supported in part by NSF Grant No.~PHY-1915005. C.B.V.~also acknowledges support from Simons Investigator Award \#376204. R.R.~acknowledges support from the National Science Foundation Graduate Research Fellowship Program under Grant No.~1839285.

\appendix

\section{Energies}
\label{app:energies}
In this appendix we derive two relations relating to the energy of Proca solitons. We begin with the definition of the charge $Q$ in \eqref{e.Qdef}:
\begin{align}
Q&=4\pi \Phi_0^2\int \dd\rho\,\rho^2f^2\left(\Omega-\alpha A \right) \\
&=\frac{4\pi\Phi^2_0M^2}{\alpha}\int \dd\rho\,\rho^2A-\frac{4\pi\Phi^2_0}{\alpha}\lim_{\rho\to\infty}\rho^2A' \,,
\end{align}
where the second line uses the $A$ equation given in~\eqref{e.Aeq}.
If $M=0$, this implies that for large $\rho$
\begin{equation}
A=\frac{\alpha\, Q}{4\pi\,\Phi^2_0\,\rho},\label{e.asymG}
\end{equation}
up to corrections that fall off faster than $1/\rho$~\cite{Lee:1988ag}. For $M\neq0$, requiring a finite $Q$ implies $A$ falls off faster than $1/\rho$ and we obtain
\beq
\frac{Q\alpha}{4\pi\Phi_0^2M^2}=\int \dd\rho\,\rho^2A~.\label{e.StuckQ}
\eeq
These results lead to a rewriting of the soliton energy given in Eq.~\eqref{e.Eint}:
\begin{align}
E/\sqrt{m_\phi^2-\omega_0^2}&=4\pi\Phi_0^2\int \dd\rho\,\rho^2\left[ \frac12f^{\prime 2}+\frac{1}{2}A^{\prime2}+f^2(\Omega-\alpha A)^2+M^2A^2-V(f,A)\right],\nonumber\\
&=4\pi\Phi_0^2\int \dd\rho\,\rho^2\left[ \frac12f^{\prime 2}+\frac{1}{2}A^{\prime2}+\frac{1}{\alpha\rho^2}(A\alpha-\Omega)\left( \rho^2A'\right)'+\frac{\Omega M^2}{\alpha}A-V(f,A)\right],
\end{align}
where in the second line we have used the $A$ equation from~\eqref{e.Aeq}. The third term can be integrated by parts; when $M=0$ this produces a nonzero boundary term determined by Eq.~\eqref{e.asymG}, when $M\neq0$ the boundary term vanishes and the fourth term can be evaluated using Eq.~\eqref{e.StuckQ}. In either case we find
\begin{align}
E = \omega\, Q-L\,.\label{e.simpEeq}
\end{align}
We can also compute that
\begin{align}
\frac{\dd L}{\dd\omega}&=\sqrt{m_\phi^2-\omega_0^2}\,4\pi\Phi_0^2\int \dd\rho\,\rho^2\left[-f'\frac{\dd f'}{\dd \Omega}+A'\frac{\dd A'}{\dd \Omega}+\frac{\dd V}{\dd \Omega} \right]\nonumber\\
&=\sqrt{m_\phi^2-\omega_0^2}\,4\pi\Phi_0^2\left\{\left.\rho^2A'\frac{\dd A}{\dd\Omega}\right|^\infty_0+\int \dd\rho\,\rho^2\left[-\frac{\dd f}{\dd\Omega}\frac{\partial V}{\partial f}-\frac{\dd A}{\dd\Omega}\frac{\partial V}{\partial A}+\frac{\dd V}{\dd\Omega}\right] \right\}\nonumber\\
&=\sqrt{m_\phi^2-\omega_0^2}\,4\pi\Phi_0^2\int \dd\rho\,\rho^2\frac{\partial V}{\partial\Omega}\nonumber\\
&= Q\,,
\end{align}
where one should use Eq.~\eqref{e.asymG} to evaluate the limit in the second line when $M=0$. This, along with Eq.~\eqref{e.simpEeq}, implies that
\beq
\frac{\dd E}{\dd\omega}=\omega\frac{\dd Q}{\dd\omega}~,
\eeq
which extends the result in~\cite{Gulamov:2013cra} to the case of nonzero $M$.

The form for the energy given in Eq.~\eqref{e.EnoU} is obtained by first noting the following: suppose we take the Lagrangian~\eqref{e.massesLag} and rescale the radial coordinate $\rho\to\chi \rho$. This yields
\begin{equation}
L=\sqrt{m_\phi^2-\omega_0^2}\, 4\pi\Phi_0^2\int \dd\rho\,\rho^2\chi\left[ -\frac12f^{\prime 2}+\frac{1}{2}A^{\prime2}+\chi^2V(f,A)\right].
\end{equation}
Consider the variation of this Lagrangian with respect to $\chi$ and then set $\chi=1$. The first contribution comes from the Lagrangian's explicit dependence on $\chi$, while the second follows from the implicit dependence through the functions $f(\rho)\to f(\rho\chi),\, A(\rho)\to A(\rho\chi)$. This second collection of terms, with $\chi$ set to one, is simply the usual variation of the Lagrangian, and so vanishes due to the equations of motion. Requiring the other term in the variation to also vanish yields the constraint
\begin{equation}
0=\int \dd\rho\,\rho^2\left[ -\frac12f^{\prime 2}+\frac{1}{2}A^{\prime2}+3V(f,A)\right].\label{e.Lconst}
\end{equation}
We can use this constraint to remove the explicit dependence on $U(f)$ from the energy in Eq.~\eqref{e.Eint}:
\begin{align}
E/\sqrt{m_\phi^2-\omega_0^2}&=4\pi\Phi_0^2\int \dd\rho\,\rho^2\left[\frac13f^{\prime 2}+\frac{2}{3}A^{\prime2} +f^2(\alpha A-\Omega)^2+M^2A^2 \right]\nonumber\\
&=4\pi\Phi_0^2\int \dd\rho\,\rho^2\left[\frac13f^{\prime 2}+\frac{2}{3}A^{\prime2}+\frac{1}{\alpha\rho^2} (\alpha A-\Omega)\left(\rho^2A' \right)'+\frac{M^2\Omega}{\alpha}\right],
\end{align}
where in the last line we have used the $A$ equation of motion. This third term is then integrated by parts to produce
\begin{align}
E/\sqrt{m_\phi^2-\omega_0^2} = \Omega Q+\frac{4\pi\Phi_0^2}{3}\int \dd\rho\,\rho^2\left(f^{\prime 2}-A^{\prime2} \right) ,
\end{align}
which is the desired result.

\bibliographystyle{utcaps_mod}
\bibliography{BIB}

\end{document}